\documentclass[onecolumn,11pt, notitlepage]{article}

\usepackage{graphicx}
\usepackage{amsmath,amsfonts,amssymb,amsthm,color}
\usepackage{authblk}

\usepackage[round]{natbib}
\usepackage{hyperref}
\hypersetup{
    colorlinks = true,
    urlcolor   = blue,
    citecolor  = black,
}

\newcommand{\RomanNumeralCaps}[1]

\usepackage[labelsep=endash,labelfont=bf,justification=justified]{caption}
\usepackage[margin=1in]{geometry}

\usepackage{float}
\usepackage{caption}
\usepackage[linesnumbered,ruled,vlined]{algorithm2e}

\title{Detecting Lagrangian coherent structures \\ from sparse and noisy trajectory data}

\author[1]{Saviz Mowlavi}
\author[2,3]{Mattia Serra\footnote{mserra@ucsd.edu}}
\author[4]{Enrico Maiorino}
\author[2,5,6]{L Mahadevan\footnote{lmahadev@g.harvard.edu}}

\affil[1]{Department of Mechanical Engineering, Massachusetts Institute of Technology, Cambridge, MA 02139, USA}
\affil[2]{School of Engineering and Applied Sciences, Harvard University, Cambridge, MA 02138, USA}
\affil[3]{Department of Physics, University of California, San Diego, CA 92093, USA}
\affil[4]{Channing Division of Network Medicine, Harvard Medical School, Boston, MA 02115, USA}
\affil[5]{Department of Organismic and Evolutionary Biology, Harvard University, Cambridge, MA 02138, USA}
\affil[6]{Department of Physics, Harvard University, Cambridge, MA 02138, USA}
  
\date{}

\begin{document}

\maketitle

\begin{abstract}
Many complex flows such as those arising from ocean plastics in geophysics or moving cells in biology are characterized by sparse and noisy trajectory datasets.
We introduce techniques for identifying Lagrangian Coherent Structures (LCSs) of hyperbolic and elliptic nature in such datasets.
Hyperbolic LCSs, which represent surfaces with maximal attraction or repulsion over a finite amount of time, are computed through a regularized least-squares approximation of the flow map gradient. Elliptic LCSs, which identify regions of coherent motion such as vortices and jets, are extracted using DBSCAN -- a popular data clustering algorithm -- combined with a systematic approach to choose parameters. We deploy these methods on various benchmark analytical flows and real-life experimental datasets ranging from oceanography to biology and show that they yield accurate results, despite sparse and noisy data. We also provide a lightweight computational implementation of these techniques as a user-friendly and straightforward Python code.
\end{abstract}

\section{Introduction}

Coherent material structures are ubiquitous at all length scales, from oceanic and atmospheric processes \citep{haller2015,serra2017uncovering,serra2020search} to biological systems \citep{serra2020}. There exists a plethora of physical and biological systems defined by the dynamics of an ensemble of particles in space and time; examples include drifter trajectories in the ocean \citep{lumpkin2007}, cell motion in living systems \citep{hogan1999}, bacteria or other constituent agents in active fluids \citep{marchetti2013,morozov2017}, and so on. In general, the coordinated motion of material parcels can be visualized by the organized patterns that tracer particles form over time \citep{merzkirch2012}. However, simply observing instantaneous particle positions and velocities provides an incomplete and deceptive picture -- for instance, a simple change to a rotating reference frame suffices to alter the observed coherent patterns. This motivated the development of the theory of Lagrangian Coherent Structures (LCSs) \citep{shadden2012,haller2015}, as well as their infinitesimally short time analogs called Objective Eulerian Coherent Structures \citep{serra2016objective,nolan2020finite},  which provide a frame-invariant framework for identifying the flow structures shaping observed patterns.

\begin{figure}
\centering
\includegraphics[width=\textwidth]{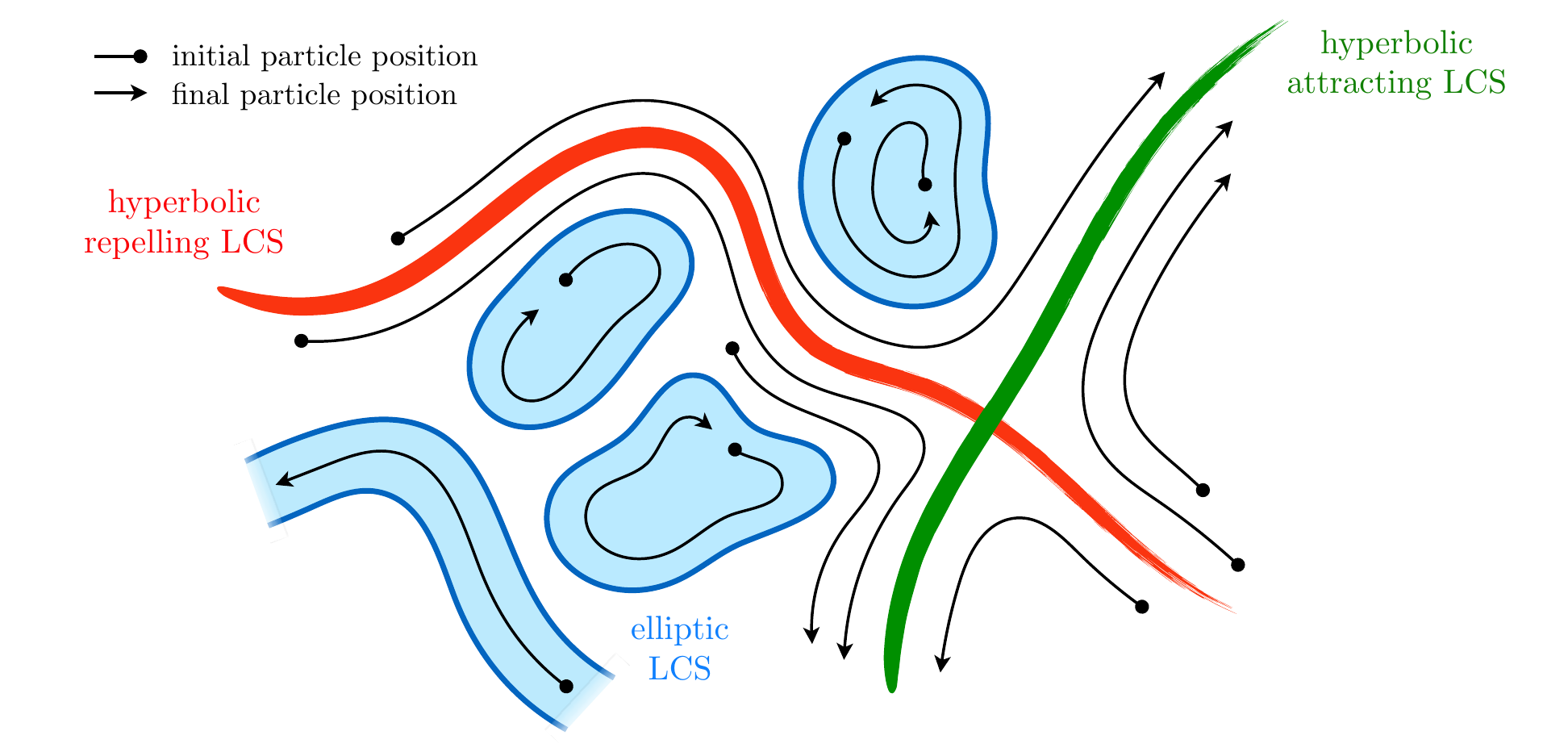}
\caption{Sketch of the different types of Lagrangian Coherent Structures (LCS) that are considered in this paper. While the flow field pictured here is time-invariant to ease visualization, these structures also apply to general time-dependent flows. Note that we include elongated structures such as jets, originally classified as parabolic LCSs \citep{haller2015}, in our definition of elliptic LCSs.}
\label{fig:CoherentStructures}
\end{figure}

The computation of LCSs requires Lagrangian particle trajectories over a finite time interval, and they may be classified into two main groups -- hyperbolic and elliptic LCSs, illustrated in Figure \ref{fig:CoherentStructures}.
Hyperbolic LCSs are surfaces along which the local separation rate between neighboring particles is maximized or minimized, leading to their binary classification as repelling or attracting LCSs, respectively. Elliptic LCSs are surfaces enclosing regions of coherent global dynamics, that is, regions inside of which particles move together over time. Although these regions often consist of vortical flow structures, we also include elongated structures such as jets, originally classified as parabolic LCSs \citep{haller2015}, in our definition of elliptic LCSs. Together, hyperbolic and elliptic LCSs provide a complete picture of coordinated motion in fluid flows and other physical or biological systems.

Early attempts to compute LCSs from flow data \citep[reviewed in][]{hadjighasem2017} required knowledge of a dense set of particle trajectories. In practice, this is only obtainable through numerical integration of a well-resolved velocity field, from e.g.~computational simulations or particle image velocimetry (PIV) of experimental data, but not from particle tracking velocimetry (PTV) or single-cell tracking, which provide sparser trajectories. Therefore, extending computational tools for LCS identification to sparse and possibly noisy data would open the door to a wider array of potential applications, leading to a better understanding of the structure and properties of a broad class of particle-based physical systems.

Hyperbolic LCSs are based on the separation rate between initially-close material parcels, making their identification using sparse and noisy data difficult. Their computation usually relies on the finite-time Lyapunov exponent (FTLE), which measures the sensitivity of final particle positions with respect to their initial positions \citep{haller2001,shadden2005}. However, the accurate calculation of this quantity requires the trajectories of a dense and regular array of initial particle positions, rendering its extension to sparse datasets nontrivial. In fact, the only existing methods for computing hyperbolic LCSs that accept irregular initial positions are the finite-time entropy (FTE) framework of \cite{froyland2012}, the set-oriented redefinition of the FTLE proposed by \cite{tallapragada2013}, and the trajectory-stretching exponent (TSE) defined in \cite{haller2021}, all of which bypass the calculation of the true FTLE. The first two methods partition the spatial domain into a set of boxes, evaluate a discrete transfer operator that quantifies the probability of particles transferring between any two boxes over the time interval of interest \citep{froyland2009}, and compute different diagnostic quantities to measure local stretching. Nevertheless, the robustness of these approaches to sparsity and noise in the data has not been established. The third method computes the TSE, a brilliant measure of local material stretching using only individual trajectories, and is therefore well-suited to sparse data despite being only quasi-objective.

Elliptic LCSs are global features that characterize the overall behavior of a connected set of material parcels. Taking advantage of the fact that they can be identified indirectly through the coherent sets that they enclose \citep{froyland2010,hadjighasem2016}, several techniques for identifying elliptic LCSs from sparse data have recently been proposed and are reviewed in \cite{hadjighasem2017}. Methods applicable to three-dimensional flows fall broadly into two categories. The first category comprises methods based on the notion of coherent set introduced in \cite{froyland2010} and \cite{froyland2013}, which are region that minimize mixing with surrounding material elements in the presence of diffusion. These sets are calculated from the transfer operator of \cite{froyland2009} using a variety of techniques \citep{ser2015,froyland2015b,williams2015,banisch2017,froyland2019}. In the second category, individual trajectories are interpreted as points in an abstract space endowed with a certain notion of distance, and various clustering tools from computer science and statistics \citep{fortunato2010,everitt2011} are used to group trajectories that are close together into separate clusters. 
Such techniques differ from one another not only by the clustering algorithm they employ, but also by their definition of distance between particle trajectories. They include the application of fuzzy clustering \citep{froyland2015c}, spectral graph partitioning \citep{hadjighasem2016,padberg2017,banisch2017,wichmann2020,vieira2020}, spectral graph drawing \citep{schlueter2017a}, and density-based clustering \citep{schneide2018,wichmann2021}. However, all of the aforementioned methods struggle to determine consistently the correct number of clusters, even in simple analytical flows such as the Bickley jet \citep{hadjighasem2017}. The number of clusters is either required as a heuristic input to the algorithm that often fails as the data gets sparse and/or noisy, or is dependent on the specific choice of parameters for the method.

Here, we introduce two techniques for computing hyperbolic and elliptic LCSs using sparse and noisy trajectory datasets that solve the aforementioned issues. Our approach to compute hyperbolic LCSs follows the widely used FTLE-based definition and relies on a local least squares fit of the flow map gradient, a tensorial quantity from which the FTLE is calculated. By feeding pairwise stretching information from all particles in a small neighborhood around the location of interest, the method is able to alleviate the effects of both sparsity and noise while retaining enough spatial resolution to resolve the ridges in the resulting FTLE field, which locate hyperbolic LCSs. Next, our procedure to identify elliptic LCSs utilizes the same clustering algorithm as in \cite{schneide2018} -- density-based spatial clustering of applications with noise (DBSCAN) -- due to its many advantages. First, it is able to tell apart trajectories belonging to coherent structures from those that do not, a feature that partition-based methods struggle to achieve \citep{froyland2019,wichmann2021}. Second, it identifies both compact structures such as vortices and elongated structures such as jets, because it assigns trajectories to clusters based on their local proximity to neighboring trajectories in the cluster rather than all of them, as in spectral clustering \citep{hadjighasem2016}. Third, its implementation is straightforward due to the many scientific libraries that it is implemented in, and it does not require a subsequent clustering step like many of the aforementioned approaches do \citep{hadjighasem2016,schlueter2017b}. Here we propose a consistent procedure for selecting the clustering parameters in the DBSCAN algorithm. Overall, we demonstrate the robustness of our methods to different systems using both analytical and experimental data, polluted by various levels of noise and sparsity. We also provide user-friendly, lightweight Python codes implementing these techniques on any dataset of particle trajectories.

The paper is organized as follows. In Section \ref{sec:Methods} we describe our methods for identifying hyperbolic and elliptic LCSs from sparse and noisy trajectory datasets. In Section \ref{sec:Results} these methods are applied to benchmark analytical flows as well as experimental datasets, and we conclude with some general remarks in Section \ref{sec:Conclusions}.

\section{Methods}
\label{sec:Methods}

We consider a discrete set of $N$ particles enclosed by a time-dependent body $\Omega(t)$ and following the trajectories $\{\mathbf{x}^i(t)\}_{i=1}^N \in \Omega(t) \subset \mathbb{R}^d$, where time $t \in [t_0,t_f]$ and $d$ is the spatial dimension of the system. In the following, we will find convenient to introduce the flow map $\mathbf{F}_{t_0}^t(\mathbf{x}_0^i) = \mathbf{x}^i(t)$, which takes the initial position $\mathbf{x}_0^i$ of particle $i$ at time $t_0$ to its current position $\mathbf{x}^i(t)$ at time $t$.

\subsection{Hyperbolic coherent structures}

Hyperbolic structures can be divided into two categories -- repelling (attracting) LCSs are surfaces along which the separation (attraction) rate between neighboring particles on either side of the surface at initial (final) time is maximized. While a number of different theories have been proposed to identify hyperbolic LCSs in fluid flows \citep[for reviews, see][]{haller2015,allshouse2015}, we propose an adaptation to sparse datasets of the original technique based on the finite-time Lyapunov exponent (FTLE) \citep{haller2001,shadden2005}, widely used due to its simplicity.

\subsubsection{Dense trajectory or velocity datasets}

We first review the definition of the FTLE field in the case where the flow map $\mathbf{F}_{t_0}^t(\mathbf{x}_0) = \mathbf{x}(t)$ is known for every initial position $\mathbf{x}_0 \in \Omega(t_0)$. In practice, this condition is met when one has access to a continuous or discretized version of the entire velocity field of the system (over space and time), which can be used to numerically integrate the trajectory $\mathbf{x}(t)$ of any particle. 

For two particles initially located at $\mathbf{x}_0$ and $\mathbf{x}_0+d\mathbf{x}_0$, the separation between them at time $t$ is
\begin{equation}
\mathbf{F}_{t_0}^t(\mathbf{x}_0+d\mathbf{x}_0)-\mathbf{F}_{t_0}^t(\mathbf{x}_0) = \nabla \mathbf{F}_{t_0}^t(\mathbf{x}_0) d\mathbf{x}_0+ O(|d\mathbf{x}_0|^2),
\end{equation} 
where $\nabla \mathbf{F}_{t_0}^t(\mathbf{x}_0)$ is the gradient of the flow map, usually computed using finite-difference techniques \citep{haller2001,shadden2012}. In the limit $|d\mathbf{x}_0| \ll 1$, the ratio $\lambda$ of initial and final distances between these two particles is
\begin{equation}
\lambda(\mathbf{x}_0;d\mathbf{x}_0) = \frac{|\mathbf{F}_{t_0}^t(\mathbf{x}_0+d\mathbf{x}_0)-\mathbf{F}_{t_0}^t(\mathbf{x}_0)|}{|d\mathbf{x}_0|} \simeq \frac{|\nabla \mathbf{F}_{t_0}^t(\mathbf{x}_0) d\mathbf{x}_0|}{|d\mathbf{x}_0|}.
\label{eq:Stretch}
\end{equation}
The largest possible stretch ratio $\lambda$ over all infinitesimal segments $d\mathbf{x}_0$ is equal to the largest singular value of the flow map gradient, which typically grows exponentially \citep{wiggins2003}. The forward-time FTLE is defined as the ratio of the exponent of this growth to the time interval,
\begin{align}
\Lambda_{t_0}^t(\mathbf{x}_0) = \frac{1}{t-t_0} \ln \left[ \max_{d\mathbf{x}_0} \lambda(\mathbf{x}_0;d\mathbf{x}_0)\right ] = \frac{1}{t-t_0} \ln \left [s_1(\mathbf{x}_0)\right ],
\label{eq:FTLE}
\end{align}
where $s_1(\mathbf{x}_0)$ denotes the largest singular value of $\nabla \mathbf{F}_{t_0}^t(\mathbf{x}_0)$. Since $\Lambda_{t_0}^t(\mathbf{x}_0)$ quantifies the rate of local material deformation, one can define the initial position of a repelling hyperbolic LCS over the time interval $[t_0,t]$ as the ridges of the scalar field $\Lambda_{t_0}^t(\mathbf{x}_0)$ (which correspond to curves in the plane and surfaces in space). Conversely, attracting hyperbolic LCSs can be thought of as repelling LCSs in backward time; therefore, the final position of an attracting LCS over $[t_0,t]$ can be defined as the ridges of the backward-time FTLE scalar field $\Lambda_{t}^{t_0}(\mathbf{x})$, where $\mathbf{x} = \mathbf{F}_{t_0}^t(\mathbf{x}_0)$.

\subsubsection{Sparse and noisy trajectory datasets}
\label{sec:DiscreteFTLE}

\begin{figure}
\centering
\includegraphics[width=\textwidth]{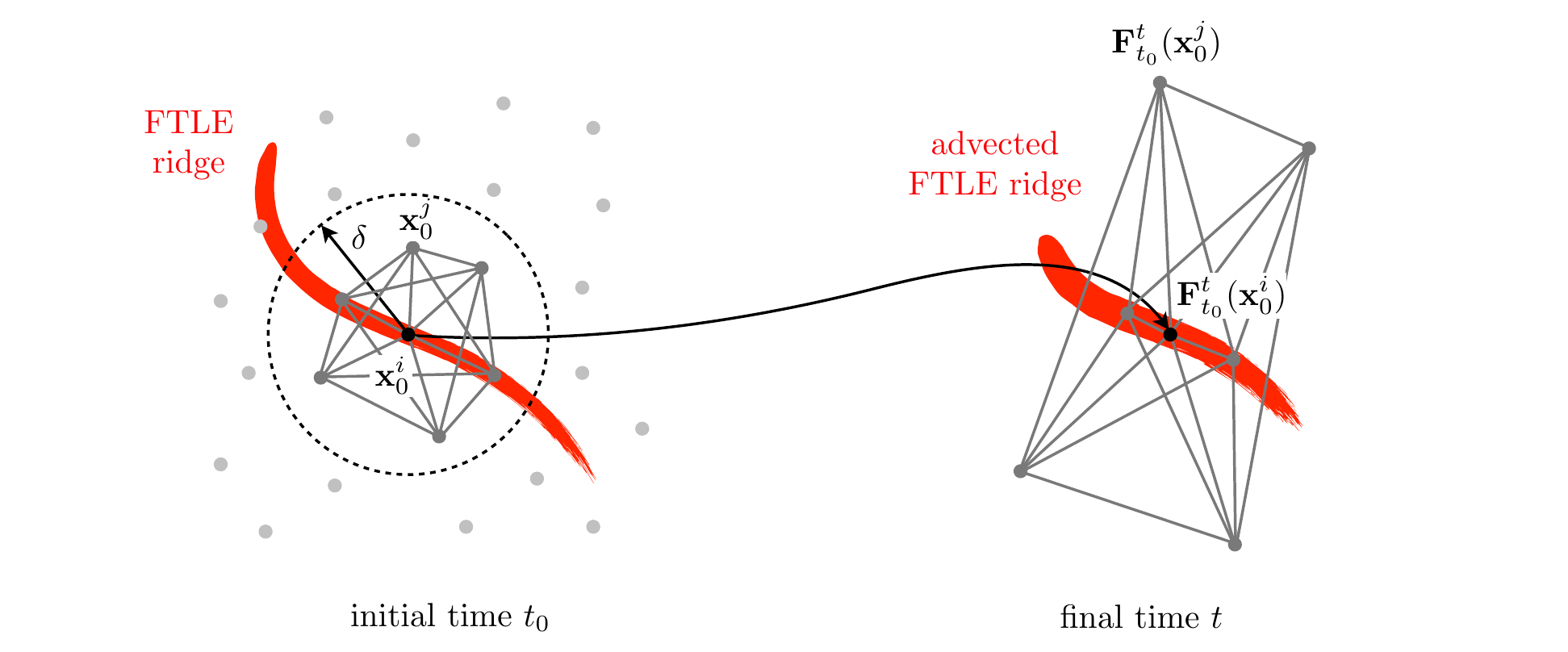}
\caption{Identification of hyperbolic repelling LCSs. The flow map gradient is approximated as the least-square fit of the tensor that maps the initial positions of segments relating $\mathbf{x}_0^i$ and its neighbors in a neighborhood of radius $\delta$ to their final positions.}
\label{fig:FTLE}
\end{figure}

The flow map gradient $\nabla \mathbf{F}_{t_0}^t$, which enters the usual definition \eqref{eq:FTLE} of the FTLE field, is inaccessible when one only has access to the trajectories of $N$ particles $\{\mathbf{x}^i(t)\}_{i=1}^N$. In this case, we introduce an approximation of the flow map gradient, $\tilde{\nabla} \mathbf{F}_{t_0}^t$, which we calculate at every particle initial position $\mathbf{x}_0^i = \mathbf{x}^i(t_0)$ as follows. First, we draw a ball of radius $\delta$ around the initial position $\mathbf{x}_0^i$ of a given particle $i$, and call $\mathcal{N}$ the set of all particles $j$, including $i$ itself, such that $|\mathbf{x}_0^j - \mathbf{x}_0^i| < \delta$, as pictured in Figure \ref{fig:FTLE}.
The approximate flow map gradient at $\mathbf{x}_0^i$ is then defined such that the relation
\begin{equation}
\mathbf{F}_{t_0}^t(\mathbf{x}_0^k)-\mathbf{F}_{t_0}^t(\mathbf{x}_0^j) \simeq \tilde{\nabla} \mathbf{F}_{t_0}^t(\mathbf{x}_0^i) [\mathbf{x}_0^k - \mathbf{x}_0^j]
\label{eq:ApproximatedFlowMapGradient}
\end{equation}
holds for all pairs $j,k \in \mathcal{N}, j \neq k$. Assuming that there are $M$ such pairs, the approximate flow map gradient can then be obtained by minimizing the square error between both sides of \eqref{eq:ApproximatedFlowMapGradient}, resulting in the least squares problem
\begin{equation}
\tilde{\nabla} \mathbf{F}_{t_0}^t(\mathbf{x}_0^i) = \arg \min_\mathbf{A} \ \frac{1}{M} \sum_{\substack{j,k \in \mathcal{N} \\ j \neq k}} \left\Vert \mathbf{A} [\mathbf{x}_0^k - \mathbf{x}_0^j] - [\mathbf{F}_{t_0}^t(\mathbf{x}_0^k)-\mathbf{F}_{t_0}^t(\mathbf{x}_0^j)] \right\Vert_2^2 + \textcolor{black}{\beta} \Vert \mathbf{A} - \mathbf{I} \Vert_F^2,
\label{eq:Minimization}
\end{equation}
where $\Vert\cdot\Vert_2$ and $\Vert\cdot\Vert_F$ denote the Euclidian and Frobenius norms, respectively, and \textcolor{black}{$\beta$} is a regularization parameter that biases $\tilde{\nabla} \mathbf{F}_{t_0}^t$ towards the identity in the absence of data, which corresponds to the classical Tikhonov regularization in inverse problems \citep{kaipio2006}. As shown in Appendix \ref{app:LeastSquares}, the solution of the above minimization problem is given by
\begin{equation}
\tilde{\nabla} \mathbf{F}_{t_0}^t(\mathbf{x}_0^i) = (\mathbf{Y} \mathbf{X}^\mathsf{T} + \beta M \mathbf{I}) (\mathbf{X} \mathbf{X}^\mathsf{T} + \beta M \mathbf{I})^{-1},
\label{eq:LeastSquareEstimate}
\end{equation}
where the matrices $\mathbf{X}, \mathbf{Y} \in \mathbb{R}^{d \times M}$ are defined as
\begin{gather}
\mathbf{X} = \big[ \dots \, | \, (\mathbf{x}_0^k - \mathbf{x}_0^j) \, | \, \dots \big], \\
\mathbf{Y} = \big[ \dots \, | \, (\mathbf{F}_{t_0}^t(\mathbf{x}_0^k)-\mathbf{F}_{t_0}^t(\mathbf{x}_0^j)) \, | \, \dots \big],
\label{eq:XYMatrices}
\end{gather}
for all pairs $j,k  \in \mathcal{N}, j \neq k$. While our approach is reminiscent of the work of \cite{lekien2010} to compute FTLEs on unstructured meshes, important differences include the specific set of particle pairs considered in the least squares problem \eqref{eq:Minimization}, and the addition of a regularization term. We now replace the continuous singular value $s_1(\mathbf{x}_0)$ in \eqref{eq:FTLE} with $\tilde{s}_1(\mathbf{x}_0^i)$, the largest singular value of the approximate flow map gradient $\tilde{\nabla} \mathbf{F}_{t_0}^t(\mathbf{x}_0^i)$ obtained from \eqref{eq:LeastSquareEstimate}, which yields the approximate forward-time FTLE at the particle position $\mathbf{x}_0^i$:
\begin{equation}
\tilde{\Lambda}_{t_0}^t(\mathbf{x}_0^i) = \frac{1}{t-t_0} \ln \left [\tilde{s}_1(\mathbf{x}_0^i)\right].
\label{eq:ApproxFTLE}
\end{equation}
Finally, the approximate initial positions of repelling LCSs over $[t_0,t_f]$ can be defined as the ridges of the forward-time discrete FTLE $\tilde{\Lambda}_{t_0}^t(\mathbf{x}_0^i)$. Conversely, the approximate final positions of attracting LCSs over $[t_0,t_f]$ can be defined as the ridges of the backward-time discrete FTLE $\tilde{\Lambda}_t^{t_0}(\mathbf{x}_t^i)$, where $\mathbf{x}_t^i = \mathbf{x}^i(t)$ are the final particle positions, and the set $\mathcal{N}$ now comprises all particles $j$, including $i$ itself, such that $|\mathbf{x}_t^j - \mathbf{x}_t^i| < \delta$.

We note that \eqref{eq:ApproxFTLE} could in principle be calculated at an arbitrary spatial location $\mathbf{x}$, instead of an initial particle position $\mathbf{x}_0^i$. Indeed, replacing $\mathcal{N}$ with the set of all particles $j$ such that $|\mathbf{x}_0^j - \mathbf{x}| < \delta$ would yield a continuous representation of the approximate FTLE field, $\tilde{\Lambda}_{t_0}^t(\mathbf{x})$. Here we simply compute the FTLE at initial particle positions $\mathbf{x}_0^i$ due to ease of implementation.

Our approximate scheme requires us to choose the parameters $\beta$ and $\delta$. Although both parameters play a role in reducing the effect of noise, they do so in different ways -- large $\beta$ values give less weight to the data-dependent term in \eqref{eq:Minimization}, while large $\delta$ values result in a larger set $\mathcal{N}$ of particles. In practice, we have found that $\delta$ is much more effective than $\beta$ at countering the effect of noise in the data.  {Thus, denoting with $r$ the average initial distance between neighboring particles, we choose a value $\beta \ll \sqrt{r}$ that only serves to regularize the least squares solution \eqref{eq:LeastSquareEstimate} in the degenerate case where $\mathbf{X} \mathbf{X}^\mathsf{T}$ is ill-conditioned. In the examples to follow, we select $\beta = 10^{-10}$. For $\delta$ we start from a value on the order of $r$, and slowly increase it until clean ridges emerge from the computed FTLE field.} This will be illustrated with examples in Sections \ref{sec:Bickley}, \ref{sec:ABC}, and \ref{sec:Chick}.

\subsection{Elliptic coherent structures}
\label{sec:EllipticLCS}

Elliptic LCSs are surfaces enclosing regions of coherent motion in which particles remain close together. Here, we consider such coherent regions to comprise both vortex-type structures and elongated ones such as jets. Similarly to hyperbolic LCSs, there exists many different techniques for identifying elliptic LCSs in fluid flows, usually through the coherent regions that they enclose \citep[for a review, see][]{hadjighasem2017}. 

As mentioned in the introduction, we focus on the class of methods that interpret individual trajectories as points in an abstract space endowed with a certain notion of \textcolor{black}{distance,} before using a clustering tool to identify coherent regions as clusters of trajectories that remain close over the considered time interval. We define the distance $d_{ij}$ between two trajectories $\mathbf{x}_i(t)$ and $\mathbf{x}_j(t)$ as the time average
\begin{equation}
d_{ij} = \frac{1}{t-t_0} \int_{t_0}^t \Vert \mathbf{x}^i(t') - \mathbf{x}^j(t') \Vert dt',
\label{eq:EdgeWeight}
\end{equation}
following \cite{hadjighasem2016}, who called it the dynamical distance. Treating trajectories like individual points, we then apply the density-based spatial clustering of applications with noise (DBSCAN) algorithm introduced by \cite{ester1996}, which was first utilized to detect coherent sets of trajectories in \cite{schneide2018}.

\begin{figure}
\centering
\includegraphics[width=\textwidth]{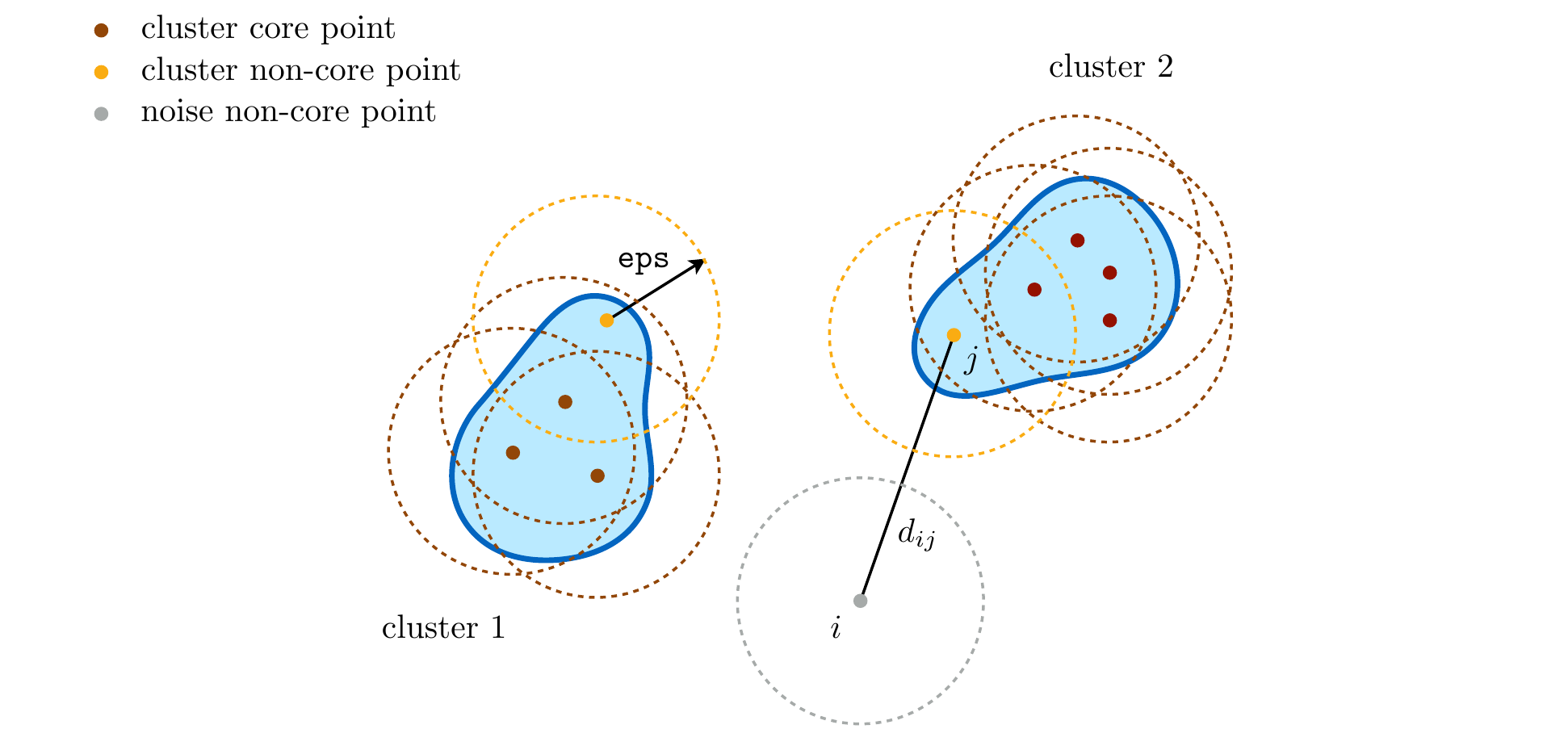}
\caption{Identification of elliptic LCSs. The DBSCAN algorithm is applied to a set of points representing individual particle trajectories and separated by the distance defined in \eqref{eq:EdgeWeight}. It then identifies elliptic LCSs as clusters of trajectories that are close to each other.}
\label{fig:DBSCAN}
\end{figure}

The DBSCAN algorithm is implemented in many scientific libraries, making it quick and easy to run, and can be abstracted as shown in Algorithm \ref{alg:DBSCAN} \citep{schubert2017}. Given a set of points representing individual particle trajectories, separated by the distance defined in \eqref{eq:EdgeWeight}, the algorithm assigns points to clusters based on their local proximity with a given number of other points in the cluster rather than all of them. This produces groups defined by a minimum density throughout rather than proximity between all member points.  {The minimum density is set by the parameters \texttt{minPts} and \texttt{eps}: a minimum number \texttt{minPts} of points must belong to the same ball of radius \texttt{eps} in order to be part of the same cluster.} As a result, not only is DBSCAN able to identify both compact and elongated structures such as vortices and jets, but it also has an inherent ability to designate points that do not belong to any cluster, lying alone in low-density regions, as noise.  The resulting clusters identify groups of particles with coherent motion and enclosed by elliptic LCSs, with noise particles in between separating them, as illustrated in Figure \ref{fig:DBSCAN}.
In practice, we use the \texttt{scikit-learn} implementation of the DBSCAN algorithm in Python \citep{pedregosa2011}.

\begin{algorithm}[h]
\DontPrintSemicolon
\SetAlgoLined
Define a neighborhood of distance \texttt{eps} around every point, and identify as \textit{core points} those that have at least \texttt{minPts} neighbors\;
Create separate clusters for each group of neighboring core points; that is, core points within a distance \texttt{eps} of each other\;
Assign each non-core point to a cluster if it is in the neighborhood of a core point; otherwise, label it as noise
\caption{DBSCAN (abstract form)}
\label{alg:DBSCAN}
\end{algorithm}

A critical aspect of any clustering method consists in the choice of the parameters, which in this case are  {\texttt{minPts} and \texttt{eps}}. Although our approach based on the DBSCAN algorithm is similar to that of \cite{schneide2018}, we provide a consistent methodology for the choice of the parameter \texttt{eps}, given a value for  \texttt{minPts}. We will describe this methodology in Section \ref{sec:Bickley}, along with its application to the Bickley jet, and verify that it produces consistent results in other systems in Sections \ref{sec:ABC} and \ref{sec:GulfMexico}.

\section{Results}
\label{sec:Results}

We now apply our algorithms for the detection of hyperbolic and elliptic LCSs to four different systems. The first two, the Bickley jet and ABC flow, have analytic velocity fields and are frequently used as benchmark problems for the detection of coherent structures \citep{hadjighasem2016,hadjighasem2017}. We use the given velocity fields to calculate trajectory data from randomly-seeded particles with artificially-introduced noise, and we compare the LCSs identified by our methods with the known ground truth. The third and fourth systems, moving cells in a developing chicken embryo and ocean tracers in the Gulf of Mexico, correspond to experimental data from developmental biology and oceanography respectively; and highlight the capability of our methods to identify structures from sparse and noisy trajectory data.

\subsection{Bickley jet}
\label{sec:Bickley}

\begin{figure}
\centering
\includegraphics[width=\textwidth]{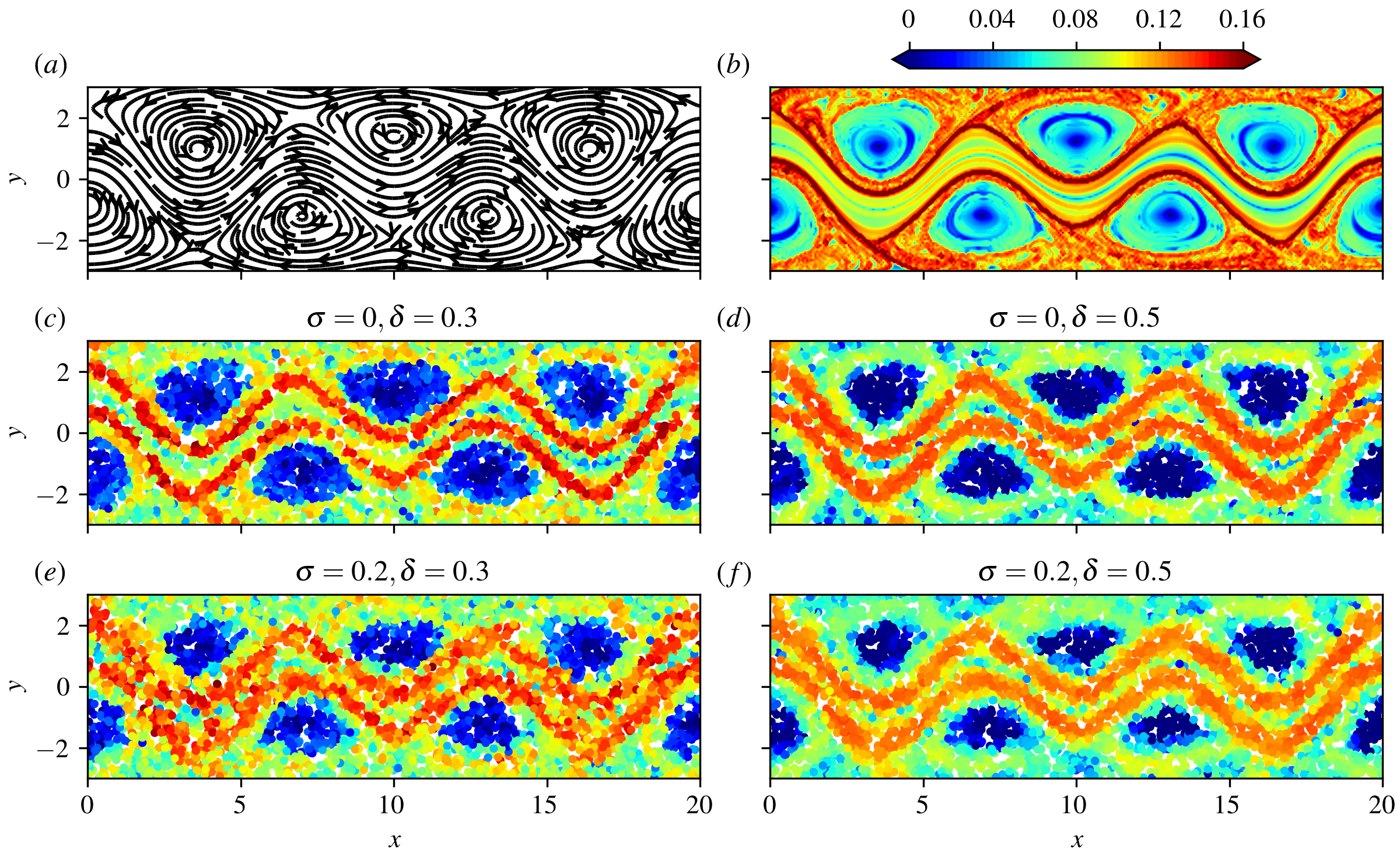}
\caption{Hyperbolic repelling LCSs in the Bickley jet. ($a$) Streamlines at $t=0$,  ($b$) exact continuous FTLE field over the time window $t \in [0,40]$, and discrete FTLE field computed over the same time window using 6000 particles with ($c$) $\sigma = 0$, $\delta = 0.3$, ($d$) $\sigma = 0$, $\delta = 0.5$, ($e$) $\sigma = 0.2$, $\delta = 0.3$, ($f$) $\sigma = 0.2$, $\delta = 0.5$. Here, $\sigma$ refers to the standard deviation of observation noise, and $\delta$ is the neighborhood radius used in the least squares fit of the flow map gradient. The colorbar is shared between plots ($b$--$f$), and values above 0.16 are shown in the same color.}
\label{fig:Bickley_FTLE}
\end{figure}

The Bickley jet is an analytical model of a two-dimensional meandering zonal jet separating counter-rotating vortices in the Earth's atmosphere \citep{del1993,rypina2007}. The time-dependent flow field is described by the streamfunction $\psi(x,y,t) = \psi_0(y) + \psi_1(x,y,t)$, where
\begin{subequations}
\begin{align}
\psi_0(y) &= c_3 y - UL \tanh (y/L), \\
\psi_1(x,y,t) &= UL \, \mathrm{sech}^2 (y/L) \sum_{n=1}^3 \epsilon_n \cos(k_n(x - \omega_n t)). 
\end{align}
\end{subequations}
In order to facilitate comparison, we use the same parameter values as in \cite{schlueter2017a}; that is, $U = 62.66 \, \mathrm{m \, s}^{-1}$, $L = 1.77 \cdot 10^6 \, \mathrm{m}$, $k_n = 2n/r_0$ where $r_0 = 6.371 \cdot 10^6 \, \mathrm{m}$ is the radius of the Earth, $\epsilon = [0.0075, 0.15, 0.3]$, $c = [0.1446, 0.205, 0.461] U$, and $\textcolor{black}{\omega} = c - c_3$. The flow is computed in the $x$-periodic domain $\Omega = [0,\pi r_0] \times [-3,3] 10^6 \, \mathrm{m}$. In the following, all distance units are in $10^6 \, \mathrm{m}$ and time units are in days. The streamlines produced at $t=0$ by these parameter values are depicted in Figure \ref{fig:Bickley_FTLE}($a$), and show the existence of two pairs of counter-rotating vortices separated by a meandering horizontal jet. The streamfunction and parameter values considered here produce the same flow field as that studied in \cite{hadjighasem2017}, but viewed in a reference frame translating at the constant speed of the vortices.

We first compute repelling LCSs in the Bickley jet over the time window $t \in [0,40]$, visualized by the ridges of the forward FTLE field at initial time. As a benchmark, Figure \ref{fig:Bickley_FTLE}($b$) displays the reference FTLE field $\Lambda_0^{40}$, computed through a finite-difference approximation of the flow map gradient \citep{shadden2012} using a set of particles initialized on a $300 \times 90$ rectangular grid. Particle trajectories were integrated from the velocity field using a 4th order Runge Kutta method with tolerance $10^{-6}$, as will be the case for all trajectory calculations in the remainder of the paper. We note that for the purpose of evaluating the flow map gradient, we let trajectories leave the initial domain $\Omega$ despite the periodicity in the $x$-direction. \textcolor{black}{Indeed, the flow map gradient relates the initial and final values of the separation vector between infinitesimally close trajectories. Since this separation vector lives in the so-called `tangent space' of $\Omega$, which is devoid of any periodicity, the flow map gradient is better approximated by evaluating the difference between unbounded trajectories.}

When working with particle trajectories from experimental data that is both sparse and associated with random initial locations, we need to modify the finite-difference methodology utilized to deduce the results in Figure \ref{fig:Bickley_FTLE}($b$). To simulate such a scenario using the Bickley jet, we compute trajectories for 6000 particles whose initial positions are randomly assigned through a spatially-uniform distribution function, and we add Gaussian noise of standard deviation $\sigma$ to each spatio-temporal measurement to simulate observation noise in real data. Figures \ref{fig:Bickley_FTLE}($c$--$f$) show the discrete FTLE field $\tilde{\Lambda}_0^{40}$ for different values of $\sigma$ and of the neighborhood radius $\delta$. 
In a manner similar to the continuous FTLE field in ($b$), these discrete FTLE fields also display the two ridges separating the meandering jet from the vortices. In the absence of noise, a smaller value of $\delta$ enables a finer delineation of the ridges. However, in the presence of noise, increasing $\delta$ leads to an effective averaging that reduces the pollution caused by the noise and yields cleaner ridges, due to the additional data used in the least squares fit of the flow map gradient at every point. {Therefore, $\delta$ controls a trade-off between the ridges' width and their pollution from noise. Because its optimal value depends on the inherent spatial resolution and noisiness of the data, $\delta$ should in practice be slowly increased until clean ridges emerge from the computed FTLE field, but not made too large that it will eventually smear the ridges.}

We turn to the computation of elliptic LCSs over $t \in [0,40]$. The FTLE in Figure \ref{fig:Bickley_FTLE}($b$) suggests the existence of such structures enclosing seven regions of the flow -- each of the six vortices as well as the meandering jet, which has been confirmed by distinct methods \citep{hadjighasem2017,vieira2020,wichmann2021}. We consider 1080 particles initialized on a rectangular grid and pollute their trajectories with Gaussian noise of standard deviation $\sigma$. We then calculate  $d_{ij}$ accounting for the periodicity of the domain in the $x$-direction, and then apply DBSCAN.
\textcolor{black}{Because the periodicity of the domain sets an upper bound on $d_{ij}$,} this makes it more difficult for the clustering algorithm to discriminate between particles belonging to different groups of coherent motion\footnote{For instance, the value of $d_{ij}$ between a particle trajectory in the meandering jet and another in one of the vortices will be much smaller than if we did not account for domain periodicity due to the much faster horizontal velocity of the jet. This smaller value would then be closer to the value of $d_{ij}$ between two particles trajectories in the same vortex.}. Some of the previous attempts at calculating elliptic LCSs using clustering algorithms did not account for such periodicity in the calculation of pairwise distances \citep{hadjighasem2016,schlueter2017a}. Finally, we note that the clustering results are largely insensitive to small perturbations in the initial positions of the particles; thus a rectangular grid was chosen simply in order to aid the visualization of the computed structures.

\begin{figure}
\centering
\includegraphics[width=\textwidth]{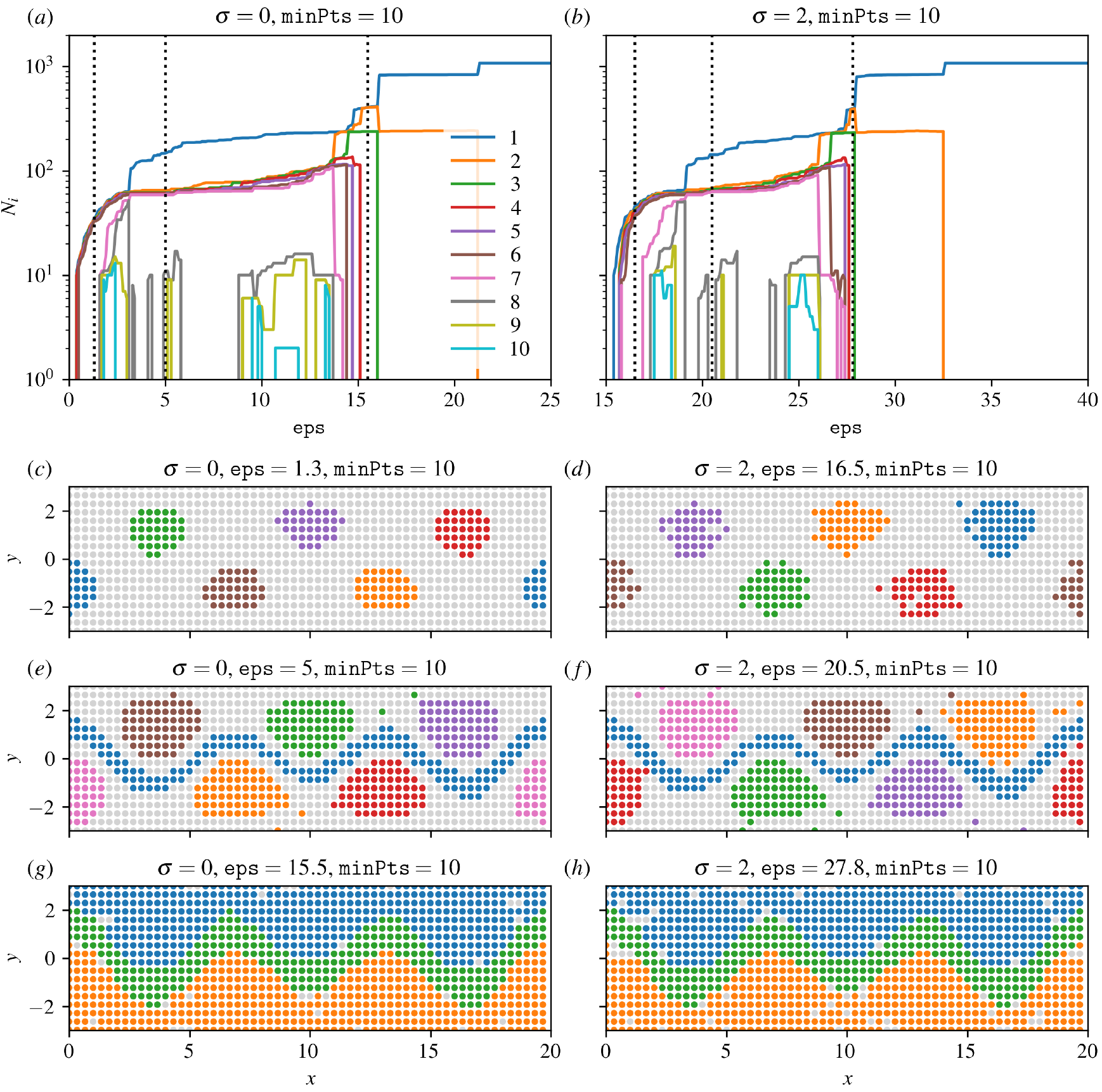}
\caption{Elliptic LCSs in the Bickley jet. ($a,b$) Number $N_i$ of particles in each of the ten largest groups identified by the DBSCAN algorithm as a function of the parameter $\texttt{eps}$ for $\texttt{minPts} = 10$. The data consists of 1080 particles trajectories advected over $t \in [0,40]$ days and artificially polluted by observation noise of standard deviation ($a$) $\sigma = 0$ and ($b$) $\sigma = 2$.  ($c$--$h$) Coherent groups identified by DBSCAN for ($c,e,g$) $\sigma = 0$, ($d,f,h$) $\sigma = 2$, and different values of $\texttt{eps}$ shown as the dotted lines in ($a$,$b$). {\href{https://www.dropbox.com/s/d9h5npthwkqc3cp/MovieBickley.mp4?dl=0}{Movie1} shows the Lagrangian evolution of the particles in ($e$).}}
\label{fig:Bickley_DBSCAN}
\end{figure}

Beginning with no observation noise, Figure \ref{fig:Bickley_DBSCAN}($a$) shows the number $N_i$ of particles in each of the ten largest groups identified by the DBSCAN algorithm as a function of the parameter $\texttt{eps}$ for $\texttt{minPts} = 10$.
The algorithm classifies all particles as noise (i.e., not part of any coherent group) for $\texttt{eps} = 0$, and lumps all particles into one unique group for large $\texttt{eps}$. This behavior is expected given that DBSCAN creates groups out of core particles with a minimum of \texttt{minPts} neighboring particles in a neighborhood of radius \texttt{eps}. As \texttt{eps} is varied in between these two extremes, we observe two categories of groups. Groups 1 to 7 contain a rather stable number of particles over a wide range of \texttt{eps}, while groups 8 to 10 repeatedly appear and disappear under minute changes of \texttt{eps}. The first category contains physically meaningful groups, while the second category contains spurious groups that we henceforth discard in the following plots. Figures \ref{fig:Bickley_DBSCAN}($c,e,g$) display the coherent groups identified by DBSCAN for the three different values of \texttt{eps} shown as the dotted lines in ($a$). A small \texttt{eps} merely identifies the vortex cores while a large \texttt{eps} detects the meandering jets but not the vortices. Thus, \texttt{eps} sets the coherence length scale of the identified structures, as \cite{wichmann2021} noticed with the OPTICS algorithm, itself a generalization of DBSCAN. We argue, however, that \textit{an appropriate choice for \texttt{eps} is one that does not induce changes in the number of physically meaningful groups as its value is slightly perturbed}, which is a widespread philosophy in clustering applications \citep{luxburg2010}. Figure \ref{fig:Bickley_DBSCAN}($a$) indicates that any \texttt{eps} in the range 3.5 to 14 generates the same number of clusters, which comprise both the meandering jet and the vortices as shown in ($e$) for $\texttt{eps} = 5$. \textcolor{black}{Thus, any value of $\texttt{eps}$ in the range 3.5 to 14 and not too close to its boundaries is a valid choice. In particular, the clustering result obtained for $\texttt{eps} = 5$} identifies all known elliptic LCSs in the Bickley jet, and therefore validates our methodology for the choice of \texttt{eps}, which can be summarized as the following steps:
\begin{enumerate}
\item Discard spurious groups that repeatedly appear and disappear under very small changes of $\texttt{eps}$ (in Figure \ref{fig:Bickley_DBSCAN}(a), groups 8 and above),
\item Identify the range of $\texttt{eps}$ in which the number of remaining physically meaningful groups does not change (in Figure \ref{fig:Bickley_DBSCAN}(a), $3.5 \le \texttt{eps} \le 14$),
\item Select a value of $\texttt{eps}$ in that range, preferably not too close to its boundaries.
\end{enumerate}
We note that the few seemingly outlier particles away from the green and orange clusters in panel ($e$) correctly belong to the assigned groups, as demonstrated in Appendix \ref{app:ApparentOutliersBickleyJet}.  {\href{https://www.dropbox.com/s/d9h5npthwkqc3cp/MovieBickley.mp4?dl=0}{Movie1} shows the Lagrangian evolution of the particles in ($e$).} 

We then consider the effect of observation noise. Figures \ref{fig:Bickley_DBSCAN}($b,d,f,h$) are the counterparts of ($a,c,e,g$), with Gaussian noise of standard deviation $\sigma = 2$ applied to each spatio-temporal data point in the computed particle trajectories. To facilitate visualization, the particles in ($d,f,h$) are displayed at their true initial location without the applied noise. Despite the intensity of the noise, the coherent groups identified by the clustering algorithm are remarkably similar with their noise-free counterparts. The main difference is that similar clustering results correspond to higher values of \texttt{eps} in the presence of noise, but our methodology itself is robust to noise.

Finally, we present in Appendix \ref{app:BickleyJetResultsLessParticles} a similar analysis as carried out in this section, using fewer particle trajectories. The results demonstrate the robustness of our algorithms for the detection of hyperbolic and elliptic coherent structures to various amounts of noise and sparsity in the data.

\subsection{ABC flow}
\label{sec:ABC}

\begin{figure}
\centering
\includegraphics[width=\textwidth]{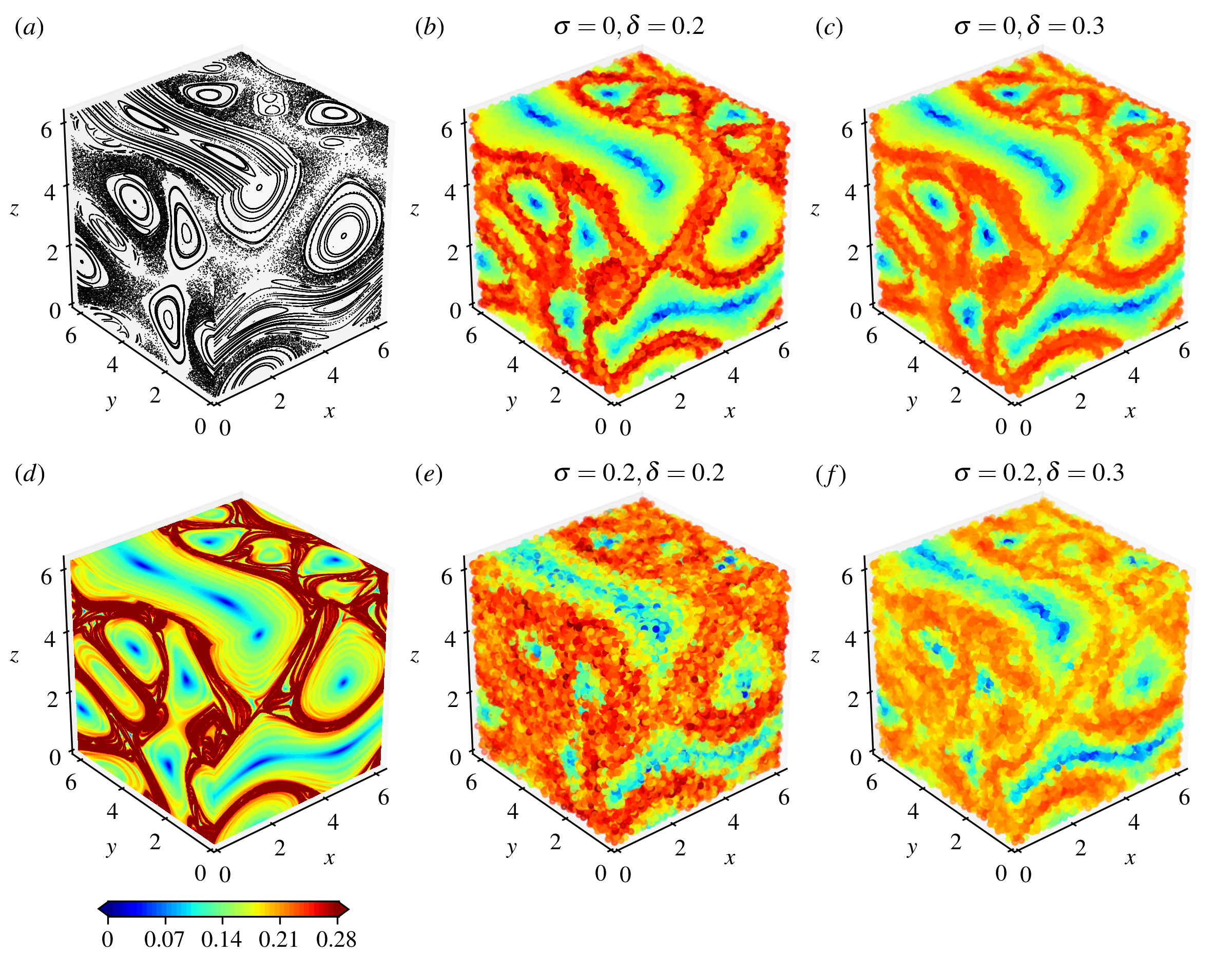}
\caption{Hyperbolic repelling LCSs in the ABC flow. ($a$) Poincare maps for integration time $t \in [0,2000]$,  ($d$) exact continuous FTLE field over the time window $t \in [0,20]$, and discrete FTLE field computed over the same time window using 200~000 particles with ($b$) $\sigma = 0$, $\delta = 0.2$, ($c$) $\sigma = 0$, $\delta = 0.3$, ($e$) $\sigma = 0.2$, $\delta = 0.2$, ($f$) $\sigma = 0.2$, $\delta = 0.3$. Here, $\sigma$ refers to the standard deviation of observation noise, and $\delta$ is the neighborhood radius used in the least squares fit of the flow map gradient. The colorbar is shared between plots ($b$--$f$), and values below 0 and above 0.28 are shown in the same color.}
\label{fig:ABC_FTLE}
\end{figure}

For our second analytical example, we consider the ABC (Arnold--Beltrami--Childress) flow, a family of exact three-dimensional solutions of Euler's equation given by the steady velocity field
\begin{subequations}
\begin{align}
\dot{x} &= A \sin z + C \cos y, \\
\dot{y} &= B \sin x + A \cos z, \\
\dot{z} &= C \sin y + B \cos x.
\end{align}%
\end{subequations}%
We employ the same parameter values $A = \sqrt{3}$, $B = \sqrt{2}$, and $C = 1$ considered in previous studies \citep{dombre1986,haller2001} and set the domain to be the three-dimensional torus $\Omega = [0,2\pi]^3$ with periodic boundary conditions \citep{dombre1986,froyland2009}. Figure \ref{fig:ABC_FTLE}($a$) shows three orthogonal Poincaré sections of 100 trajectories initialized on each face of the cube and advected over $t \in [0,2000]$.
These Poincaré sections confirm the existence of six regions of coherent motion that have been discovered previously \citep{dombre1986,budivsic2012,hadjighasem2016}.

We first compute the repelling LCSs in the ABC flow over the time window $t \in [0,20]$, visualized by the ridges of the forward FTLE field. Figure \ref{fig:ABC_FTLE}($d$) displays the reference FTLE field $\Lambda_0^{20}$ at initial time, computed by finite-differencing the flow map obtained through advection of a dense rectangular initial lattice of $200 \times 200 \times 200$ particles. Similar to our treatment of the Bickley jet, we let the trajectories leave the initial domain $\Omega$ when evaluating the flow map gradient. As expected, the FTLE ridges bound the regions of coherent motion revealed by the Poincaré sections in Figure \ref{fig:ABC_FTLE}($a$). We then simulate sparse experimental data by computing the trajectories of 200~000 particles randomly initialized in $\Omega$ following a spatially-uniform distribution, and we apply a Gaussian noise of standard deviation $\sigma$ to each resulting spatio-temporal measurement. Figures \ref{fig:ABC_FTLE}($b$,$c$,$e$,$f$) show the discrete FTLE field $\tilde{\Lambda}_0^{20}$ for different values of $\sigma$ and of the neighborhood radius $\delta$. 
As we have observed with the Bickley jet, a smaller value of $\delta$ enables a finer delineation of the ridges in the absence of noise. However, under the presence of noise, increasing $\delta$ reduces the pollution caused by the noise and leads to cleaner ridges. These results confirm the important role played by the parameter $\delta$, in the case of three-dimensional data as well.

\begin{figure}
\centering
\includegraphics[width=\textwidth]{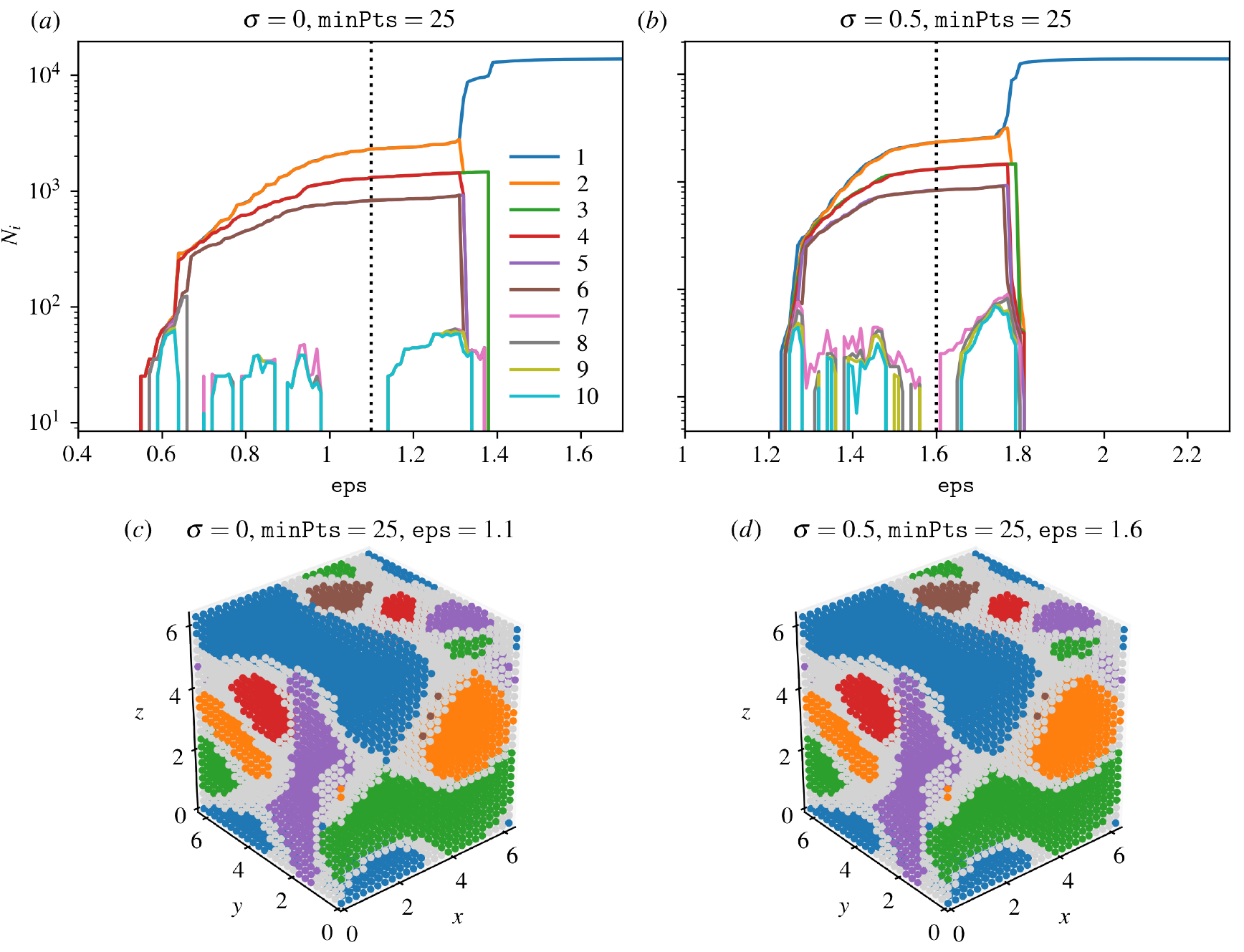}
\caption{Elliptic LCSs in the ABC flow. ($a,b$) Number $N_i$ of particles in each of the ten largest groups identified by the DBSCAN algorithm as a function of the parameter $\texttt{eps}$, for $\texttt{minPts} = 25$. The data consists of 15~625 particles trajectories advected over $t \in [0,20]$ and artificially polluted by observation noise of standard deviation ($a$) $\sigma = 0$ and ($b$) $\sigma = 0.5$.  ($c,d$) Coherent groups identified by DBSCAN for ($c$) $\sigma = 0$ and ($d$) $\sigma = 0.5$, and the value of $\texttt{eps}$ shown as the dotted line in ($a$,$b$).
 {\href{https://www.dropbox.com/s/ydbaxvxuy2splm6/MovieABC.mp4?dl=0}{Movie2} shows the Lagrangian evolution of the particles in ($c$).}}
\label{fig:ABC_DBSCAN}
\end{figure}

We then turn to the computation of elliptic LCSs over $t \in [0,20]$, which should reveal the six coherent vortices that are seen in Figure \ref{fig:ABC_FTLE}($a$). We compute trajectories for a set of 15~625 particles initialized on a rectangular lattice, to which we add Gaussian noise of standard deviation $\sigma$. We then apply the clustering approach described in Section \ref{sec:EllipticLCS} to uncover the elliptic LCSs. Beginning with the case $\sigma = 0$, Figure \ref{fig:ABC_DBSCAN}($a$) shows the number $N_i$ of particles in each of the ten largest groups identified by the DBSCAN algorithm as a function of the parameter $\texttt{eps}$ for $\texttt{minPts} = 25$.
Following the strategy that we outlined in Section \ref{sec:Bickley} for the choice of the parameter \texttt{eps}, we first discard groups 7 to 10 (and above) as they repeatedly appear and disappear under minute changes of \texttt{eps}. The number of remaining physically meaningful groups and their size remains stable for \texttt{eps} in the range 1 to 1.3. We thus pick $\texttt{eps} = 1.1$ and plot the corresponding clusters in Figure \ref{fig:ABC_DBSCAN}($c$). These identify each of the six coherent vortices that appear in Figure \ref{fig:ABC_FTLE}($a$), providing a validation of our methodology in the case of three-dimensional chaotic flows.  {\href{https://www.dropbox.com/s/ydbaxvxuy2splm6/MovieABC.mp4?dl=0}{Movie2} shows the Lagrangian evolution of the particles in Figure \ref{fig:ABC_DBSCAN}($c$),}  {highlighting how trajectories belonging to the coherent groups behave very differently from those in the incoherent group, despite starting initially very close.}

Thanks to the ability of the DBSCAN algorithm to scale up to a large number of particles, our approach does not require sparsification of the weighted graph as opposed to the spectral clustering methodology of \cite{hadjighasem2016}. For the number of particles we consider, the DBSCAN algorithm runs in a mere 5 seconds once the edge weights \eqref{eq:EdgeWeight} are computed. Further, contrary to \cite{hadjighasem2016}, we account for the periodicity of the domain when calculating the pairwise distances in \eqref{eq:EdgeWeight}, which makes it more difficult for the clustering algorithm to discriminate between particles belonging to different groups of coherent motion. Finally, the corresponding results with Gaussian noise of standard deviation $\sigma = 0.5$ are shown in Figures \ref{fig:ABC_DBSCAN}($b$,$d$). The particles in ($d$) are displayed at their true initial noise-free location, to facilitate visualization. Once again, the remarkable similarity between the clusters obtained with and without noise exemplifies the robustness of both the algorithm and the parameter selection methodology.

\subsection{Chicken embryo}
\label{sec:Chick}

For our third example, we consider the global deformation and flow of tissue in a developing chicken embryo. The experimental data, acquired by \cite{rozbicki2015}, consists of a velocity field tracking the coordinated motion of thousands of cells during a key embryonic phase known as gastrulation. This phase is characterized by the formation of the primitive streak, a structure that marks the onset of formation of multiple tissue layers and the establishment of the overall body plan during the early stages of development \citep{stern2004}. The velocity field was obtained by imaging a live chicken embryo at regular intervals of time using light-sheet microscopy (LSM), and feeding the resulting high-resolution images of moving cells to a particle-image velocimetry (PIV) algorithm. This process was carried out over a period of 12 hours, spanning the whole formation of the primitive streak.

\begin{figure}
\centering
\includegraphics[width=\textwidth]{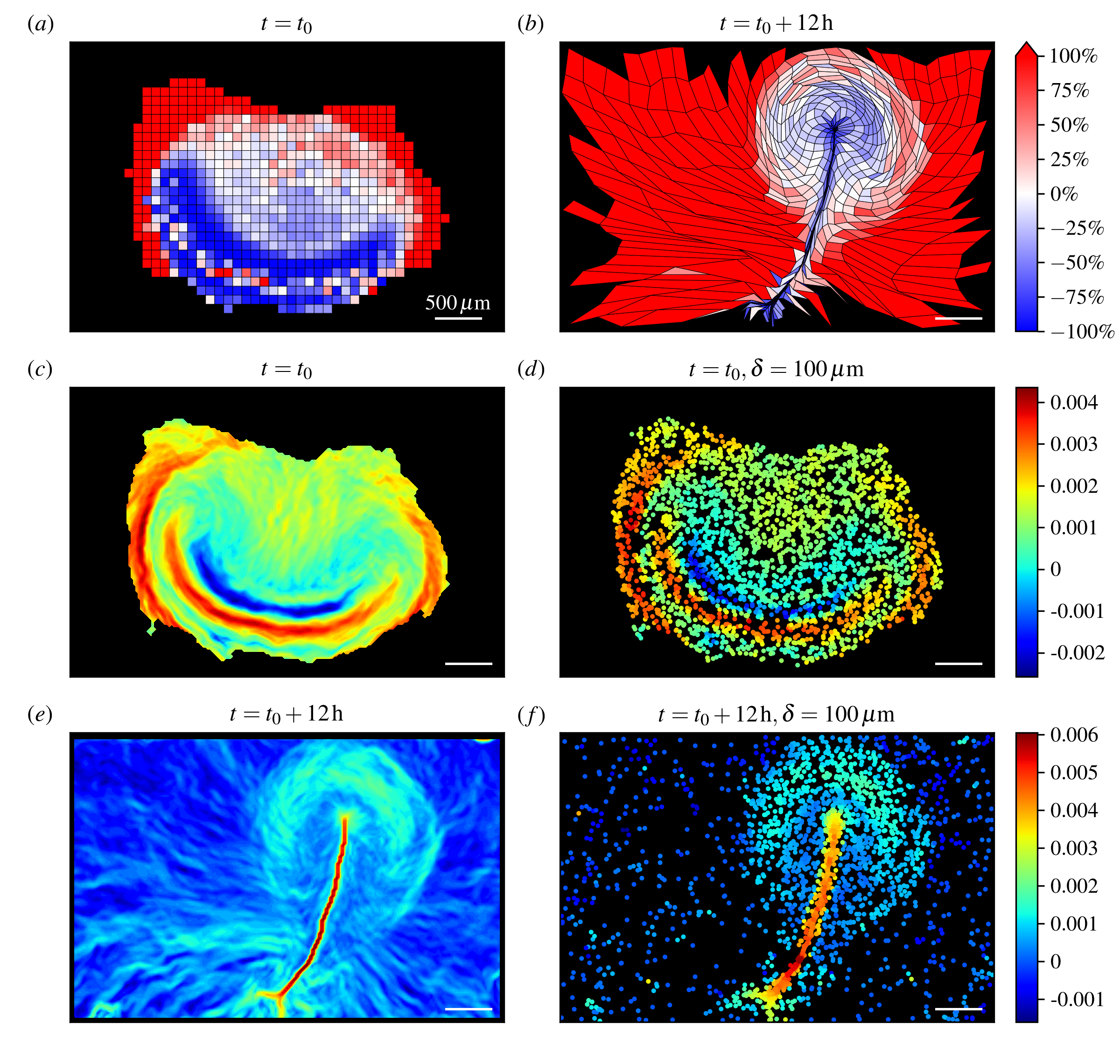}
\caption{Hyperbolic LCSs in the coordinated flow of cells belonging to a developing chick embryo, captured through light-sheet microscopy \citep{rozbicki2015}. ($a,b$) Deformation of an initially-rectangular grid advected by the flow of cells over 12 hours. Each quadrilateral is colored according to its percent volume change between initial and final times. ($c$) Exact continuous forward-time FTLE field at initial time and ($d$) its discrete counterpart, calculated from 6000 random uniformly distributed particles at initial time and advected with the flow of cells. ($e$) Exact continuous backward-time FTLE field at final time and ($f$) its discrete counterpart, calculated from the same randomly distributed particles as in ($d$). The parameter value $\epsilon = 100 \, \mu \mathrm{m}$ is used in the calculation of the discrete FTLEs in ($d$,$f$). The white scale bar corresponds to $500 \, \mu \mathrm{m}$ in all plots, and the  colorbar is shared between ($a$,$b$), ($c$,$d$), and ($e$,$f$), respectively. FTLE values are in min$^{-1}$.}
\label{fig:Chick_FTLE}
\end{figure}

We begin by visualizing the tissue deformation during formation of the primitive streak. To this effect, we advect the square tissue domains shown at initial time in Figure \ref{fig:Chick_FTLE}($a$) under the PIV-derived velocity field, resulting in the deformed domains shown in Figure \ref{fig:Chick_FTLE}($b$) at final time. 
The domains are colored according to their percent volume change over the course of the 12 hours, with blue indicating contraction and red indicating expansion. The thin elongated vertical structure in ($b$) is the primitive streak, which is formed by contraction of the deep blue domains towards it while moving away from the horizontal line of isolated red domains in ($a$) \citep{rozbicki2015}. Such deformation underlies the existence of a horizontal repelling LCS at initial time, and a vertical attracting LCS at final time \citep{serra2020}. An additional circular repelling LCSs separates the extra-embryonic region (outer red region) from the embryonic one ($a,b$).

The existence of these LCSs is confirmed in Figures \ref{fig:Chick_FTLE}($c$,$e$), in which we employ the standard finite-difference approximation of the flow map gradient to compute ($a$) the forward FTLE at initial time, and ($b$) the backward FTLE at final time. One circular and one horizontal forward FTLE ridges mark hyperbolic repelling LCSs at the initial time. The circular ridge separates embryonic and extra-embryonic areas. At the same time, particles on either side of the horizontal ridge are drawn towards the anterior and the posterior of an attracting LCS, revealed by the backward FTLE ridge. 

A continuous velocity field might not always be available, so we now consider the case where one only has discrete trajectory data for a small subset of cells. We simulate such a scenario by randomly initializing 6000 cells and advecting them under the PIV-derived velocity field. Using this single set of trajectory data, we utilize our discrete approach described in Section \ref{sec:DiscreteFTLE} with $\epsilon = 100 \, \mu \mathrm{m}$ to compute the discrete forward FTLE shown in Figure \ref{fig:Chick_FTLE}($d$) at initial time, and the discrete backward FTLE shown in Figure \ref{fig:Chick_FTLE}($f$) at final time. The discrete results agree extremely well with their continuous counterparts. Most importantly, the ridges delineating the repelling and attracting LCSs are still clearly visible, despite the sparsity of the discrete dataset. Such agreement validates the applicability of our methodology for the computation of hyperbolic LCSs to sparse and noisy real-life experimental datasets.

\subsection{Gulf of Mexico}
\label{sec:GulfMexico}

In our last example, we analyze the trajectories of a set of 152 acoustically-tracked isobaric floats in the Gulf of Mexico \citep{hamilton2016}. The floats were ballasted for depth $1500 \, \mathrm{m}$ and deployed over a two-year period starting in 2011, with each float recording its position over a duration varying between a week and 1.5 year. Based on a probabilistic Markov chain analysis \citep{froyland2014} of these trajectories, \cite{miron2019} found that the deep circulation dynamics consist of two weakly-interacting provinces of near-equal area, in each of which drifters had a residence time of 3.5 to 4.5 years. 

\begin{figure}
\centering
\includegraphics[width=\textwidth]{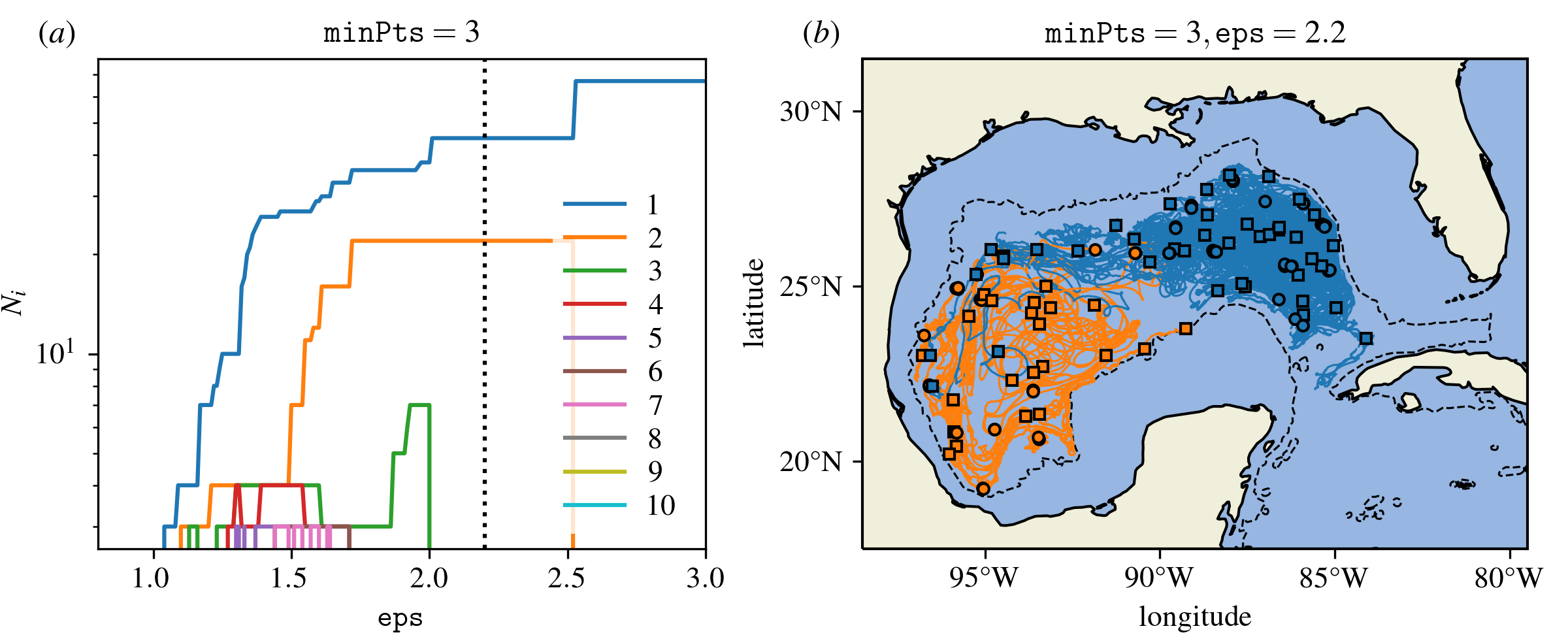}
\caption{Elliptic LCSs in the deep Gulf of Mexico, using trajectories produced over 610 days by 35 acoustically-tracked isobaric floats at depth $1500 \, \mathrm{m}$. ($a$) Number $N_i$ of floats in each of the ten largest groups identified by the DBSCAN algorithm as a function of $\texttt{eps}$, for $\texttt{minPts} = 3$.  ($b$) The two coherent groups identified by DBSCAN for $\texttt{eps} = 2.2^\circ$ are shown in blue and orange, in terms of the initial (squares) and final (circles) positions of the floats as well as their trajectories (lines). The black dotted line indicates the $1500 \, \mathrm{m}$ isobath.}
\label{fig:Ocean_DBSCAN}
\end{figure}

From the 152 float trajectories, we first select 35 of them that overlap during the same 610 days. Using the pairwise distance definition \eqref{eq:EdgeWeight}, we then apply the DBSCAN algorithm with $\texttt{minPts} = 3$. Figure \ref{fig:Ocean_DBSCAN}($a$) shows the number $N_i$ of particles in each of the ten largest resulting groups.
Following our previous methodology, we discard groups 3 to 10 (and beyond) as they repeatedly appear and disappear, or change their size by multiple factors, under minute changes of \texttt{eps}. The size of the remaining two physically meaningful groups remains constant for \texttt{eps} in the range $2^\circ$ to $2.5^\circ$. We thus pick $\texttt{eps} = 2.2^\circ$ and plot the corresponding clusters in Figure \ref{fig:Ocean_DBSCAN}($b$) by coloring the trajectories (lines), initial (squares) and final (circles) positions of the floats according to their cluster membership. We recover the two equally-sized regions of coherent motion that were identified in \cite{miron2019}. Notably, this was achieved at much lower human effort and computational cost, since the latter study requires the construction of a spatial grid over which the matrix of transition probabilities is calculated. By contrast, our approach is meshless, straightforward to implement and takes less than a second of computational time in this particular case.\\

\section{Conclusions}
\label{sec:Conclusions}

We have introduced objective computational techniques for detecting hyperbolic and elliptic Lagrangian Coherent Structures (LCSs) in complex systems based solely on knowledge of sparse and noisy particle trajectories. These techniques were validated using two benchmark problems defined by analytical velocity fields: the two-dimensional Bickley jet and the chaotic three-dimensional ABC flow. Our methodology produced accurate results at a low computational cost in both cases and exhibited good robustness to measurement noise. We then applied our techniques to two experimental datasets: the global flow of cells in a developing chicken embryo and the trajectories of ballasted floats in the Gulf of Mexico, identifying the key LCSs organizing these complex flows. 
The complexity, computational requirements and sensitivity of most existing techniques for LCSs detection have prevented their broad use in engineering, physical and biological problems. To this end, we have provided a simple, lightweight and user-friendly Python code that implements our methods on any dataset of particle trajectories.

The present work opens the door to several avenues of future research. In our approach to detect hyperbolic structures, the least-squares problem \eqref{eq:Minimization} gives equal weights to all particle pairs contained within a ball of a given radius around the point of interest. Weighting each pair according to its distance from the point of interest may improve the accuracy of the fit. 
Likewise, the clustering results of DBSCAN are heavily dependent on the definition \eqref{eq:EdgeWeight} of the distance between two trajectories, which currently measures the time average of the instantaneous pairwise distance. Taking inspiration from \cite{schlueter2017a}, one can use metrics such as the temporal standard deviation of the pairwise distance or alternative synchronization measures. 
Finally, we envisage extending our methods to account for incomplete and or newly added trajectories. Examples include ocean drifters that break down or temporarily lose GPS tracking, cells dividing into two, and dying cells. 

\section*{Acknowledgments}
We acknowledge Manli Chuai and Cornelis J. Weijer for providing the chicken embryo dataset.
M.S. acknowledges partial support from the Schmidt Science Fellowship and the Postdoc Mobility Fellowship from the Swiss National Foundation. L.M. thanks the NSF-Simons Center for Mathematical and Statistical Analysis of Biology Award 1764269, NIH 1R01HD097068, the Simons Foundation, and the Henri Seydoux Fund for partial support. 

\section*{Code accessibility}
Python codes implementing the methods described are shared in an online repository at \url{https://github.com/smowlavi/CoherentStructures} and can readily be applied to any experimental or computational dataset of particle trajectories. 

\appendix

\section{Solution of the least squares problem}
\label{app:LeastSquares}

Here, we describe how to solve the least-squares problem defined by \eqref{eq:Minimization}, which includes Tikhonov regularization \citep{kaipio2006}. With the matrices $\mathbf{X}, \mathbf{Y} \in \mathbb{R}^{2 \times m}$ introduced in \eqref{eq:XYMatrices}, one can rewrite \eqref{eq:Minimization} in matrix form as follows:
\begin{equation}
\tilde{\nabla} \mathbf{F}_{t_0}^t(\mathbf{x}_0^i) = \arg \min_\mathbf{A} J(\mathbf{A}) \equiv \arg \min_\mathbf{A} \ \frac{1}{M} \left\Vert \mathbf{A} \mathbf{X} - \mathbf{Y} \right\Vert_F^2 + \beta \Vert \mathbf{A} - \mathbf{I} \Vert_F^2.
\label{eq:MinimizationMatrix}
\end{equation}
Due to the convexity of the objective function $J(\mathbf{A})$, its global minimum is obtained when its gradient with respect to $\mathbf{A}$ vanishes. To calculate the gradient, we utilize the definition of the Frobenius norm to expand each term in the objective function $J(\mathbf{A})$ as
\begin{gather}
\left\Vert \mathbf{A} \mathbf{X} - \mathbf{Y} \right\Vert_F^2 = \text{tr}[ (\mathbf{A} \mathbf{X} - \mathbf{Y})^\mathsf{T} (\mathbf{A} \mathbf{X} - \mathbf{Y}) ] = \text{tr}[ \mathbf{X}^\mathsf{T} \mathbf{A}^\mathsf{T} \mathbf{A} \mathbf{X} - \mathbf{X}^\mathsf{T} \mathbf{A}^\mathsf{T} \mathbf{Y} - \mathbf{Y}^\mathsf{T} \mathbf{A} \mathbf{X} + \mathbf{Y}^\mathsf{T} \mathbf{Y} ], \\
\left\Vert \mathbf{A} - \mathbf{I} \right\Vert_F^2 = \text{tr}[ (\mathbf{A} - \mathbf{I})^\mathsf{T} (\mathbf{A} - \mathbf{I}) ] = \text{tr}[ \mathbf{A}^\mathsf{T} \mathbf{A} - \mathbf{A}^\mathsf{T} - \mathbf{A} +  \mathbf{I} ].
\end{gather}
Applying standard matrix calculus identities, the gradient of the objective function $J(\mathbf{A})$ with respect to matrix $\mathbf{A}$ is then given by
\begin{equation}
\frac{\partial J(\mathbf{A})}{\partial \mathbf{A}} = \frac{2}{M} \mathbf{A} \mathbf{X} \mathbf{X}^\mathsf{T} - \frac{2}{M} \mathbf{A} \mathbf{Y} \mathbf{X}^\mathsf{T} + 2 \lambda \mathbf{A} - 2 \beta \mathbf{I}.
\end{equation}
Setting the above derivative to zero and solving for $\mathbf{A}$, one finally obtains the solution given in \eqref{eq:LeastSquareEstimate}.

\section{Apparent outliers in the Bickley jet}
\label{app:ApparentOutliersBickleyJet}

\begin{figure}
\centering
\includegraphics[width=\textwidth]{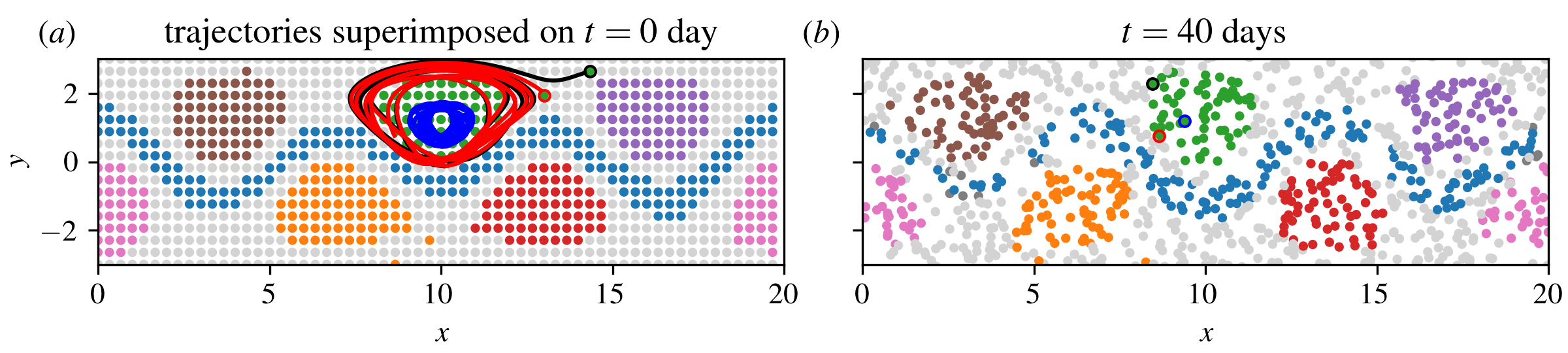}
\caption{Fate of two seemingly outlier particles in the green Bickley jet cluster obtained in Figure \ref{fig:Bickley_DBSCAN}(e) using $\texttt{minPts} = 10$, $\texttt{eps} = 5$ and $\sigma = 0$. ($a$) Initial positions (red and black circles) and trajectories (red and black lines) of these two apparent outliers superimposed with the position at $t = 0$ day of the entire clusters obtained in Figure \ref{fig:Bickley_DBSCAN}(e). We also show the trajectory (blue line) of a particle that is initially located near the middle of the green cluster. ($b$) Final positions (red and black circles) of the same two apparent outliers superimposed with the clusters displayed at $t = 40$ days, together with the final position (blue circle) of the particle that started near the middle of the green cluster.}
\label{fig:Bickley_DBSCAN_app}
\end{figure}

We analyse the trajectories of two particles labeled as belonging to the green cluster in Figure \ref{fig:Bickley_DBSCAN}(e) for the Bickley jet with $\texttt{minPts} = 10$, $\texttt{eps} = 5$ and $\sigma = 0$, despite being located away from the rest of the cluster at initial time. The red and black lines in Figure \ref{fig:Bickley_DBSCAN_app}(a) show the trajectories of these two particles with their initial positions indicated by circles, together with the trajectory, in blue, of a third particle that is initially located near the middle of the green cluster.
We also display the entire clusters obtained in Figure \ref{fig:Bickley_DBSCAN}(e) at $t = 0$ day. The trajectories reveal that the two apparent outlier particles in fact rapidly converge toward the green cluster, justifying their membership in the latter. Figure \ref{fig:Bickley_DBSCAN_app}(b) depicts the final positions at $t = 40$ days of these two outliers as the red and black circles, together with the final position of the particle that started near the middle of the green cluster as the blue circle. The entire clusters are also displayed at $t = 40$ days, showing that all three particles are located in the interior of the green cluster. Thus, the DBSCAN algorithm is correct in assigning these two apparent outliers to the green cluster. A similar analysis shows that the two apparent outliers of the orange cluster rapidly converge towards the latter, justifying their labeling as well.

\section{Bickley jet results with varying number of tracer particles}
\label{app:BickleyJetResultsLessParticles}

\begin{figure}
\centering
\includegraphics[width=\textwidth]{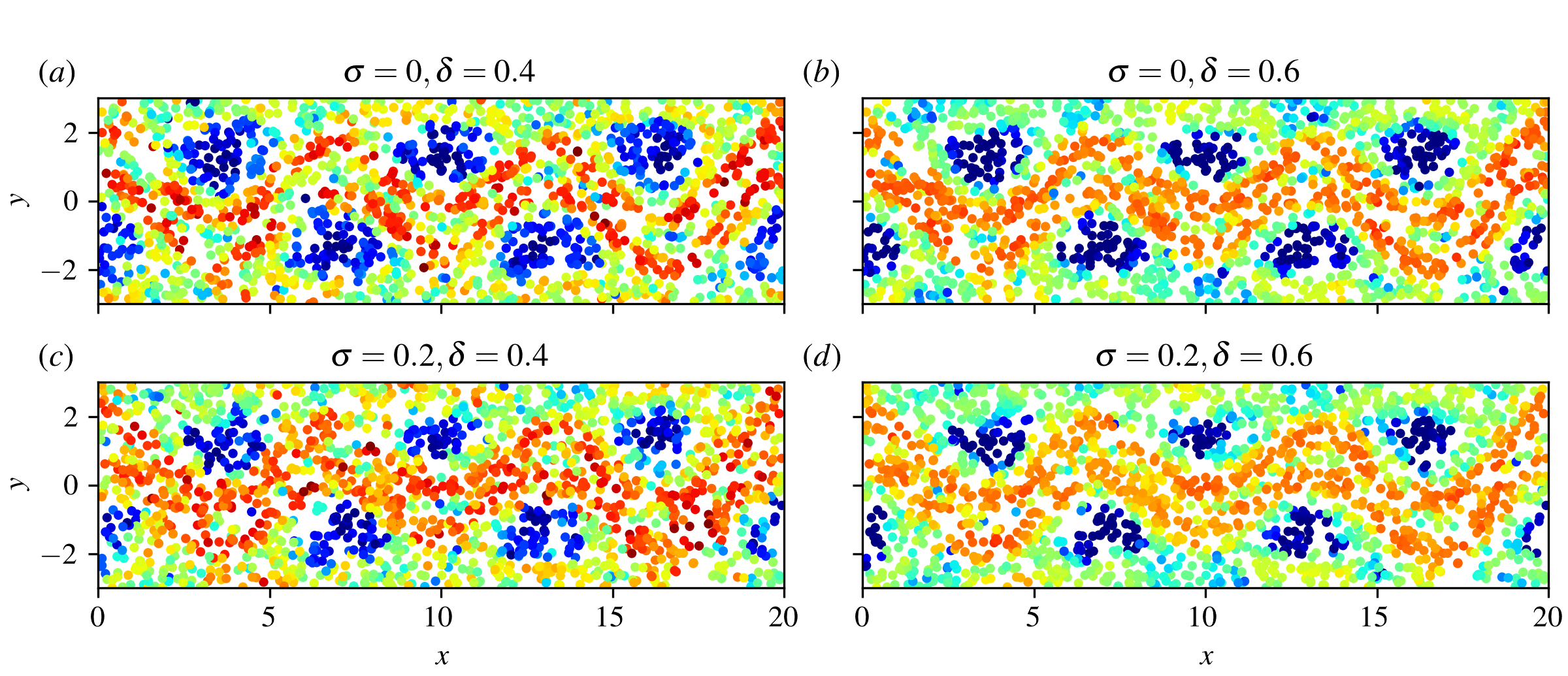}
\caption{Hyperbolic repelling LCSs in the Bickley jet with fewer particles. Discrete FTLE field computed over the time window $t \in [0,40]$ using 2000 particles with ($a$) $\sigma = 0$, $\delta = 0.4$, ($b$) $\sigma = 0$, $\delta = 0.6$, ($c$) $\sigma = 0.2$, $\delta = 0.4$, ($d$) $\sigma = 0.2$, $\delta = 0.6$. Here, $\sigma$ refers to the standard deviation of observation noise, and $\delta$ is the neighborhood radius used in the least squares fit of the flow map gradient. The colorbar is the same as in Figure \ref{fig:Bickley_FTLE}.}
\label{fig:Bickley_FTLE_2000}
\end{figure}

\begin{figure}
\centering
\includegraphics[width=\textwidth]{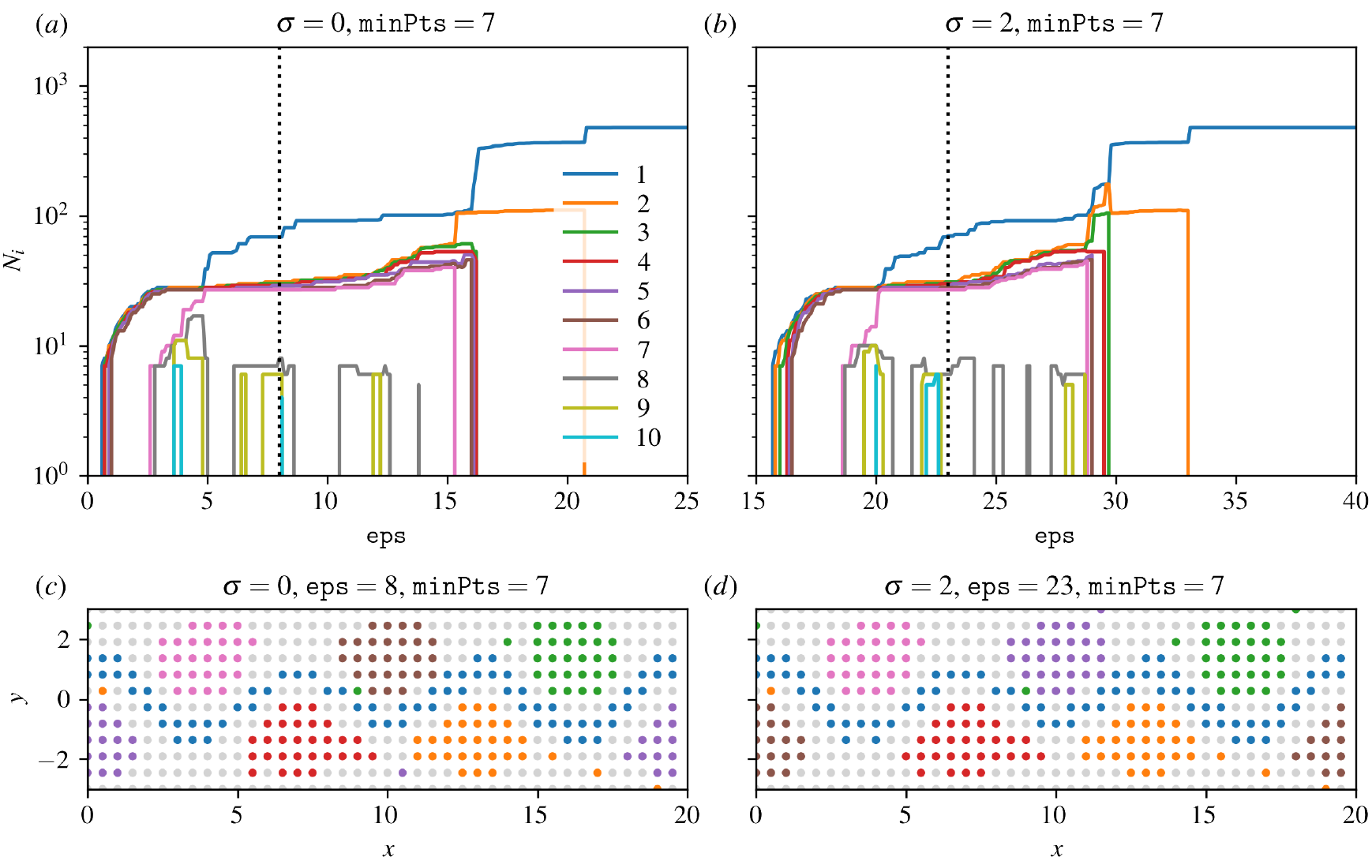}
\caption{Elliptic LCSs in the Bickley jet with fewer particles. ($a,b$) Number $N_i$ of particles in each of the ten largest groups identified by the DBSCAN algorithm as a function of the parameter $\texttt{eps}$ for $\texttt{minPts} = 7$. The data consists of 480 particles trajectories advected over $t \in [0,40]$ days and artificially polluted by observation noise of standard deviation ($a$) $\sigma = 0$ and ($b$) $\sigma = 2$.  ($c$,$d$) Coherent groups identified by DBSCAN for ($c$) $\sigma = 0$ and ($d$) $\sigma = 2$, and the value of $\texttt{eps}$ shown as the dotted line in ($a$,$b$).}
\label{fig:Bickley_DBSCAN_480}
\end{figure}

In this appendix, we demonstrate the ability of our methods to handle various levels of sparsity by revisiting the Bickley jet example investigated in Section \ref{sec:Bickley}, but using fewer particles this time. Starting with repelling LCSs, we compute trajectories for 2000 particles with random initial positions and we apply a Gaussian noise of standard deviation $\sigma$ to each measurement, as we did before. Figures \ref{fig:Bickley_FTLE_2000}($a$--$d$) show the discrete FTLE field computed using the least squares technique described in Section \ref{sec:DiscreteFTLE} for different values of $\sigma$ and of the neighborhood radius $\delta$: ($a$) $\sigma = 0$, $\delta = 0.4$, ($b$) $\sigma = 0$, $\delta = 0.6$, ($c$) $\sigma = 0.2$, $\delta = 0.4$, ($d$) $\sigma = 0.2$, $\delta = 0.6$.
Although the ridges of the FTLE field are coarser than in the presence of more data, we can still delineate the two that separate the meandering jet from the vortices. As we have observed before, increasing the parameter $\epsilon$ leads to cleaner ridges, especially in the presence of measurement noise. Next, we apply the clustering approach described in Section \ref{sec:EllipticLCS} to compute elliptic LCSs using a set of 480 particle trajectories initialized on a grid and polluted by Gaussian noise of standard deviation $\sigma$. Similar to Section \ref{sec:Bickley}, we account for the periodicity of the domain in the $x$-direction when calculating the pairwise Euclidian distance entering \eqref{eq:EdgeWeight}. Figures \ref{fig:Bickley_DBSCAN_480}($a$,$b$) show the number $N_i$ of particles in each of the ten largest groups identified by the DBSCAN algorithm as a function of the parameter $\texttt{eps}$ with $\texttt{minPts} = 7$ and for ($a$) $\sigma = 0$ and ($b$) $\sigma = 2$.
The lower value that we use here for $\texttt{minPts}$ compared with Section \ref{sec:Bickley} reflects the increased sparsity of the particle trajectories. Nevertheless, the results in Figures \ref{fig:Bickley_DBSCAN_480}($a$,$b$) are remarkably similar to their counterparts in Figures \ref{fig:Bickley_DBSCAN}($a$,$b$). Following the methodology outlined in Section \ref{sec:Bickley}, we discard spurious groups and select $\texttt{eps}$ in a range where the number and size of the remaining, physically-meaningful groups remains stable. The corresponding clusters are plotted in Figures \ref{fig:Bickley_DBSCAN_480}($c$,$d$), and they identify both the meandering jet and the vortices that were obtained in Section \ref{sec:Bickley}. Altogether, the results presented in this appendix illustrate the robustness of our detection algorithms to the simultaneous presence of noise and sparsity in the data.

\bibliographystyle{abbrvnat}
\bibliography{bibliography}

\begin{thebibliography}{54}
\providecommand{\natexlab}[1]{#1}
\providecommand{\url}[1]{\texttt{#1}}
\expandafter\ifx\csname urlstyle\endcsname\relax
  \providecommand{\doi}[1]{doi: #1}\else
  \providecommand{\doi}{doi: \begingroup \urlstyle{rm}\Url}\fi

\bibitem[Allshouse and Peacock(2015)]{allshouse2015}
M.~R. Allshouse and T.~Peacock.
\newblock Lagrangian based methods for coherent structure detection.
\newblock \emph{Chaos: An Interdisciplinary Journal of Nonlinear Science},
  25\penalty0 (9):\penalty0 097617, 2015.

\bibitem[Banisch and Koltai(2017)]{banisch2017}
R.~Banisch and P.~Koltai.
\newblock Understanding the geometry of transport: Diffusion maps for
  lagrangian trajectory data unravel coherent sets.
\newblock \emph{Chaos: An Interdisciplinary Journal of Nonlinear Science},
  27\penalty0 (3):\penalty0 035804, 2017.

\bibitem[Budi{\v{s}}i{\'c} and Mezi{\'c}(2012)]{budivsic2012}
M.~Budi{\v{s}}i{\'c} and I.~Mezi{\'c}.
\newblock Geometry of the ergodic quotient reveals coherent structures in
  flows.
\newblock \emph{Physica D: Nonlinear Phenomena}, 241\penalty0 (15):\penalty0
  1255--1269, 2012.

\bibitem[del Castillo-Negrete and Morrison(1993)]{del1993}
D.~del Castillo-Negrete and P.~Morrison.
\newblock Chaotic transport by rossby waves in shear flow.
\newblock \emph{Physics of Fluids A: Fluid Dynamics}, 5\penalty0 (4):\penalty0
  948--965, 1993.

\bibitem[Dombre et~al.(1986)Dombre, Frisch, Greene, H{\'e}non, Mehr, and
  Soward]{dombre1986}
T.~Dombre, U.~Frisch, J.~M. Greene, M.~H{\'e}non, A.~Mehr, and A.~M. Soward.
\newblock Chaotic streamlines in the abc flows.
\newblock \emph{Journal of Fluid Mechanics}, 167:\penalty0 353--391, 1986.

\bibitem[Ester et~al.(1996)Ester, Kriegel, Sander, Xu, et~al.]{ester1996}
M.~Ester, H.-P. Kriegel, J.~Sander, X.~Xu, et~al.
\newblock A density-based algorithm for discovering clusters in large spatial
  databases with noise.
\newblock In \emph{Kdd}, volume~96, pages 226--231, 1996.

\bibitem[Everitt et~al.(2011)Everitt, Landau, Leese, and Stahl]{everitt2011}
B.~S. Everitt, S.~Landau, M.~Leese, and D.~Stahl.
\newblock \emph{Cluster analysis}.
\newblock John Wiley, 2011.

\bibitem[Fortunato(2010)]{fortunato2010}
S.~Fortunato.
\newblock Community detection in graphs.
\newblock \emph{Physics reports}, 486\penalty0 (3-5):\penalty0 75--174, 2010.

\bibitem[Froyland(2013)]{froyland2013}
G.~Froyland.
\newblock An analytic framework for identifying finite-time coherent sets in
  time-dependent dynamical systems.
\newblock \emph{Physica D: Nonlinear Phenomena}, 250:\penalty0 1--19, 2013.

\bibitem[Froyland and Junge(2015)]{froyland2015b}
G.~Froyland and O.~Junge.
\newblock On fast computation of finite-time coherent sets using radial basis
  functions.
\newblock \emph{Chaos: An Interdisciplinary Journal of Nonlinear Science},
  25\penalty0 (8):\penalty0 087409, 2015.

\bibitem[Froyland and Padberg(2009)]{froyland2009}
G.~Froyland and K.~Padberg.
\newblock Almost-invariant sets and invariant manifolds—connecting
  probabilistic and geometric descriptions of coherent structures in flows.
\newblock \emph{Physica D: Nonlinear Phenomena}, 238\penalty0 (16):\penalty0
  1507--1523, 2009.

\bibitem[Froyland and Padberg-Gehle(2012)]{froyland2012}
G.~Froyland and K.~Padberg-Gehle.
\newblock Finite-time entropy: A probabilistic approach for measuring nonlinear
  stretching.
\newblock \emph{Physica D: Nonlinear Phenomena}, 241\penalty0 (19):\penalty0
  1612--1628, 2012.

\bibitem[Froyland and Padberg-Gehle(2015)]{froyland2015c}
G.~Froyland and K.~Padberg-Gehle.
\newblock A rough-and-ready cluster-based approach for extracting finite-time
  coherent sets from sparse and incomplete trajectory data.
\newblock \emph{Chaos: An Interdisciplinary Journal of Nonlinear Science},
  25\penalty0 (8):\penalty0 087406, 2015.

\bibitem[Froyland et~al.(2010)Froyland, Santitissadeekorn, and
  Monahan]{froyland2010}
G.~Froyland, N.~Santitissadeekorn, and A.~Monahan.
\newblock Transport in time-dependent dynamical systems: Finite-time coherent
  sets.
\newblock \emph{Chaos: An Interdisciplinary Journal of Nonlinear Science},
  20\penalty0 (4):\penalty0 043116, 2010.

\bibitem[Froyland et~al.(2014)Froyland, Stuart, and van Sebille]{froyland2014}
G.~Froyland, R.~M. Stuart, and E.~van Sebille.
\newblock How well-connected is the surface of the global ocean?
\newblock \emph{Chaos: An Interdisciplinary Journal of Nonlinear Science},
  24\penalty0 (3):\penalty0 033126, 2014.

\bibitem[Froyland et~al.(2019)Froyland, Rock, and Sakellariou]{froyland2019}
G.~Froyland, C.~P. Rock, and K.~Sakellariou.
\newblock Sparse eigenbasis approximation: Multiple feature extraction across
  spatiotemporal scales with application to coherent set identification.
\newblock \emph{Communications in Nonlinear Science and Numerical Simulation},
  77:\penalty0 81--107, 2019.

\bibitem[Hadjighasem et~al.(2016)Hadjighasem, Karrasch, Teramoto, and
  Haller]{hadjighasem2016}
A.~Hadjighasem, D.~Karrasch, H.~Teramoto, and G.~Haller.
\newblock Spectral-clustering approach to lagrangian vortex detection.
\newblock \emph{Physical Review E}, 93\penalty0 (6):\penalty0 063107, 2016.

\bibitem[Hadjighasem et~al.(2017)Hadjighasem, Farazmand, Blazevski, Froyland,
  and Haller]{hadjighasem2017}
A.~Hadjighasem, M.~Farazmand, D.~Blazevski, G.~Froyland, and G.~Haller.
\newblock A critical comparison of lagrangian methods for coherent structure
  detection.
\newblock \emph{Chaos: An Interdisciplinary Journal of Nonlinear Science},
  27\penalty0 (5):\penalty0 053104, 2017.

\bibitem[Haller(2001)]{haller2001}
G.~Haller.
\newblock Distinguished material surfaces and coherent structures in
  three-dimensional fluid flows.
\newblock \emph{Physica D: Nonlinear Phenomena}, 149\penalty0 (4):\penalty0
  248--277, 2001.

\bibitem[Haller(2015)]{haller2015}
G.~Haller.
\newblock Lagrangian coherent structures.
\newblock \emph{Annual Review of Fluid Mechanics}, 47:\penalty0 137--162, 2015.

\bibitem[Haller et~al.(2021)Haller, Aksamit, and Encinas-Bartos]{haller2021}
G.~Haller, N.~Aksamit, and A.~P. Encinas-Bartos.
\newblock Quasi-objective coherent structure diagnostics from single
  trajectories<? a3b2 show [feature]?>.
\newblock \emph{Chaos: An Interdisciplinary Journal of Nonlinear Science},
  31\penalty0 (4):\penalty0 043131, 2021.

\bibitem[Hamilton et~al.(2016)Hamilton, Bower, Furey, Leben, and
  P{\'e}rez-Brunius]{hamilton2016}
P.~Hamilton, A.~Bower, H.~Furey, R.~Leben, and P.~P{\'e}rez-Brunius.
\newblock Deep circulation in the gulf of mexico: A lagrangian study. ocs study
  boem 2016-081, 289 pp.
\newblock Technical report, Bureau of Ocean Energy Management, 2016.
\newblock OCS Study BOEM 2016-081.

\bibitem[Hogan(1999)]{hogan1999}
B.~L. Hogan.
\newblock Morphogenesis.
\newblock \emph{Cell}, 96\penalty0 (2):\penalty0 225--233, 1999.

\bibitem[Kaipio and Somersalo(2006)]{kaipio2006}
J.~Kaipio and E.~Somersalo.
\newblock \emph{Statistical and computational inverse problems}, volume 160.
\newblock Springer Science \& Business Media, 2006.

\bibitem[Lekien and Ross(2010)]{lekien2010}
F.~Lekien and S.~D. Ross.
\newblock The computation of finite-time lyapunov exponents on unstructured
  meshes and for non-euclidean manifolds.
\newblock \emph{Chaos: An Interdisciplinary Journal of Nonlinear Science},
  20\penalty0 (1):\penalty0 017505, 2010.

\bibitem[Lumpkin and Pazos(2007)]{lumpkin2007}
R.~Lumpkin and M.~Pazos.
\newblock Measuring surface currents with surface velocity program drifters:
  the instrument, its data, and some recent results.
\newblock \emph{Lagrangian analysis and prediction of coastal and ocean
  dynamics}, 39:\penalty0 67, 2007.

\bibitem[Marchetti et~al.(2013)Marchetti, Joanny, Ramaswamy, Liverpool, Prost,
  Rao, and Simha]{marchetti2013}
M.~C. Marchetti, J.-F. Joanny, S.~Ramaswamy, T.~B. Liverpool, J.~Prost, M.~Rao,
  and R.~A. Simha.
\newblock Hydrodynamics of soft active matter.
\newblock \emph{Reviews of Modern Physics}, 85\penalty0 (3):\penalty0 1143,
  2013.

\bibitem[Merzkirch(2012)]{merzkirch2012}
W.~Merzkirch.
\newblock \emph{Flow visualization}.
\newblock Elsevier, 2012.

\bibitem[Miron et~al.(2019)Miron, Beron-Vera, Olascoaga, Froyland,
  P{\'e}rez-Brunius, and Sheinbaum]{miron2019}
P.~Miron, F.~J. Beron-Vera, M.~J. Olascoaga, G.~Froyland, P.~P{\'e}rez-Brunius,
  and J.~Sheinbaum.
\newblock Lagrangian geography of the deep gulf of mexico.
\newblock \emph{Journal of Physical Oceanography}, 49\penalty0 (1):\penalty0
  269--290, 2019.

\bibitem[Morozov(2017)]{morozov2017}
A.~Morozov.
\newblock From chaos to order in active fluids.
\newblock \emph{Science}, 355\penalty0 (6331):\penalty0 1262--1263, 2017.

\bibitem[Nolan et~al.(2020)Nolan, Serra, and Ross]{nolan2020finite}
P.~J. Nolan, M.~Serra, and S.~D. Ross.
\newblock Finite-time lyapunov exponents in the instantaneous limit and
  material transport.
\newblock \emph{Nonlinear Dynamics}, 100\penalty0 (4):\penalty0 3825--3852,
  2020.

\bibitem[Padberg-Gehle and Schneide(2017)]{padberg2017}
K.~Padberg-Gehle and C.~Schneide.
\newblock Network-based study of lagrangian transport and mixing.
\newblock \emph{Nonlinear Processes in Geophysics}, 24\penalty0 (4):\penalty0
  661--671, 2017.

\bibitem[Pedregosa et~al.(2011)Pedregosa, Varoquaux, Gramfort, Michel, Thirion,
  Grisel, Blondel, Prettenhofer, Weiss, Dubourg, et~al.]{pedregosa2011}
F.~Pedregosa, G.~Varoquaux, A.~Gramfort, V.~Michel, B.~Thirion, O.~Grisel,
  M.~Blondel, P.~Prettenhofer, R.~Weiss, V.~Dubourg, et~al.
\newblock Scikit-learn: Machine learning in python.
\newblock \emph{the Journal of machine Learning research}, 12:\penalty0
  2825--2830, 2011.

\bibitem[Rozbicki et~al.(2015)Rozbicki, Chuai, Karjalainen, Song, Sang, Martin,
  Kn{\"o}lker, MacDonald, and Weijer]{rozbicki2015}
E.~Rozbicki, M.~Chuai, A.~I. Karjalainen, F.~Song, H.~M. Sang, R.~Martin, H.-J.
  Kn{\"o}lker, M.~P. MacDonald, and C.~J. Weijer.
\newblock Myosin-ii-mediated cell shape changes and cell intercalation
  contribute to primitive streak formation.
\newblock \emph{Nature cell biology}, 17\penalty0 (4):\penalty0 397--408, 2015.

\bibitem[Rypina et~al.(2007)Rypina, Brown, Beron-Vera, Ko{\c{c}}ak, Olascoaga,
  and Udovydchenkov]{rypina2007}
I.~Rypina, M.~G. Brown, F.~J. Beron-Vera, H.~Ko{\c{c}}ak, M.~J. Olascoaga, and
  I.~Udovydchenkov.
\newblock On the lagrangian dynamics of atmospheric zonal jets and the
  permeability of the stratospheric polar vortex.
\newblock \emph{Journal of the Atmospheric Sciences}, 64\penalty0
  (10):\penalty0 3595--3610, 2007.

\bibitem[Schlueter-Kuck and Dabiri(2017{\natexlab{a}})]{schlueter2017a}
K.~L. Schlueter-Kuck and J.~O. Dabiri.
\newblock Coherent structure colouring: identification of coherent structures
  from sparse data using graph theory.
\newblock \emph{Journal of Fluid Mechanics}, 811:\penalty0 468--486,
  2017{\natexlab{a}}.

\bibitem[Schlueter-Kuck and Dabiri(2017{\natexlab{b}})]{schlueter2017b}
K.~L. Schlueter-Kuck and J.~O. Dabiri.
\newblock Identification of individual coherent sets associated with flow
  trajectories using coherent structure coloring.
\newblock \emph{Chaos: An Interdisciplinary Journal of Nonlinear Science},
  27\penalty0 (9):\penalty0 091101, 2017{\natexlab{b}}.

\bibitem[Schneide et~al.(2018)Schneide, Pandey, Padberg-Gehle, and
  Schumacher]{schneide2018}
C.~Schneide, A.~Pandey, K.~Padberg-Gehle, and J.~Schumacher.
\newblock Probing turbulent superstructures in rayleigh-b{\'e}nard convection
  by lagrangian trajectory clusters.
\newblock \emph{Physical Review Fluids}, 3\penalty0 (11):\penalty0 113501,
  2018.

\bibitem[Schubert et~al.(2017)Schubert, Sander, Ester, Kriegel, and
  Xu]{schubert2017}
E.~Schubert, J.~Sander, M.~Ester, H.~P. Kriegel, and X.~Xu.
\newblock Dbscan revisited, revisited: why and how you should (still) use
  dbscan.
\newblock \emph{ACM Transactions on Database Systems (TODS)}, 42\penalty0
  (3):\penalty0 1--21, 2017.

\bibitem[Ser-Giacomi et~al.(2015)Ser-Giacomi, Rossi, L{\'o}pez, and
  Hern{\'a}ndez-Garc{\'\i}a]{ser2015}
E.~Ser-Giacomi, V.~Rossi, C.~L{\'o}pez, and E.~Hern{\'a}ndez-Garc{\'\i}a.
\newblock Flow networks: A characterization of geophysical fluid transport.
\newblock \emph{Chaos: An Interdisciplinary Journal of Nonlinear Science},
  25\penalty0 (3):\penalty0 036404, 2015.

\bibitem[Serra and Haller(2016)]{serra2016objective}
M.~Serra and G.~Haller.
\newblock Objective eulerian coherent structures.
\newblock \emph{Chaos: An Interdisciplinary Journal of Nonlinear Science},
  26\penalty0 (5):\penalty0 053110, 2016.

\bibitem[Serra et~al.(2017)Serra, Sathe, Beron-Vera, and
  Haller]{serra2017uncovering}
M.~Serra, P.~Sathe, F.~Beron-Vera, and G.~Haller.
\newblock Uncovering the edge of the polar vortex.
\newblock \emph{Journal of the Atmospheric Sciences}, 74\penalty0
  (11):\penalty0 3871--3885, 2017.

\bibitem[Serra et~al.(2020{\natexlab{a}})Serra, Sathe, Rypina, Kirincich, Ross,
  Lermusiaux, Allen, Peacock, and Haller]{serra2020search}
M.~Serra, P.~Sathe, I.~Rypina, A.~Kirincich, S.~D. Ross, P.~Lermusiaux,
  A.~Allen, T.~Peacock, and G.~Haller.
\newblock Search and rescue at sea aided by hidden flow structures.
\newblock \emph{Nature communications}, 11\penalty0 (1):\penalty0 1--7,
  2020{\natexlab{a}}.

\bibitem[Serra et~al.(2020{\natexlab{b}})Serra, Streichan, Chuai, Weijer, and
  Mahadevan]{serra2020}
M.~Serra, S.~Streichan, M.~Chuai, C.~J. Weijer, and L.~Mahadevan.
\newblock Dynamic morphoskeletons in development.
\newblock \emph{Proceedings of the National Academy of Sciences}, 117\penalty0
  (21):\penalty0 11444--11449, 2020{\natexlab{b}}.

\bibitem[Shadden(2012)]{shadden2012}
S.~C. Shadden.
\newblock Lagrangian coherent structures.
\newblock \emph{Transport and Mixing in Laminar Flows: From Microfluidics to
  Oceanic Currents}, pages 59--89, 2012.

\bibitem[Shadden et~al.(2005)Shadden, Lekien, and Marsden]{shadden2005}
S.~C. Shadden, F.~Lekien, and J.~E. Marsden.
\newblock Definition and properties of lagrangian coherent structures from
  finite-time lyapunov exponents in two-dimensional aperiodic flows.
\newblock \emph{Physica D: Nonlinear Phenomena}, 212\penalty0 (3-4):\penalty0
  271--304, 2005.

\bibitem[Stern(2004)]{stern2004}
C.~D. Stern.
\newblock \emph{Gastrulation: from cells to embryo}.
\newblock CSHL Press, 2004.

\bibitem[Tallapragada and Ross(2013)]{tallapragada2013}
P.~Tallapragada and S.~D. Ross.
\newblock A set oriented definition of finite-time lyapunov exponents and
  coherent sets.
\newblock \emph{Communications in Nonlinear Science and Numerical Simulation},
  18\penalty0 (5):\penalty0 1106--1126, 2013.

\bibitem[Vieira et~al.(2020)Vieira, Rypina, and Allshouse]{vieira2020}
G.~S. Vieira, I.~I. Rypina, and M.~R. Allshouse.
\newblock Uncertainty quantification of trajectory clustering applied to ocean
  ensemble forecasts.
\newblock \emph{Fluids}, 5\penalty0 (4):\penalty0 184, 2020.

\bibitem[von Luxburg(2010)]{luxburg2010}
U.~von Luxburg.
\newblock Clustering stability: An overview.
\newblock \emph{Foundations and Trends in Machine Learning}, 2\penalty0
  (3):\penalty0 235--274, 2010.

\bibitem[Wichmann et~al.(2020)Wichmann, Kehl, Dijkstra, and van
  Sebille]{wichmann2020}
D.~Wichmann, C.~Kehl, H.~A. Dijkstra, and E.~van Sebille.
\newblock Detecting flow features in scarce trajectory data using networks
  derived from symbolic itineraries: an application to surface drifters in the
  north atlantic.
\newblock \emph{Nonlinear Processes in Geophysics}, 27\penalty0 (4):\penalty0
  501--518, 2020.

\bibitem[Wichmann et~al.(2021)Wichmann, Kehl, Dijkstra, and van
  Sebille]{wichmann2021}
D.~Wichmann, C.~Kehl, H.~A. Dijkstra, and E.~van Sebille.
\newblock Ordering of trajectories reveals hierarchical finite-time coherent
  sets in lagrangian particle data: detecting agulhas rings in the south
  atlantic ocean.
\newblock \emph{Nonlinear Processes in Geophysics}, 28\penalty0 (1):\penalty0
  43--59, 2021.

\bibitem[Wiggins(2003)]{wiggins2003}
S.~Wiggins.
\newblock \emph{Introduction to applied nonlinear dynamical systems and chaos},
  volume~2.
\newblock Springer Science \& Business Media, 2003.

\bibitem[Williams et~al.(2015)Williams, Rypina, and Rowley]{williams2015}
M.~O. Williams, I.~I. Rypina, and C.~W. Rowley.
\newblock Identifying finite-time coherent sets from limited quantities of
  lagrangian data.
\newblock \emph{Chaos: An Interdisciplinary Journal of Nonlinear Science},
  25\penalty0 (8):\penalty0 087408, 2015.

\end{thebibliography}

\end{document}